\documentclass[a4paper,11pt]{article}
\pdfoutput=1 

\usepackage{jheppub} 
\usepackage{amsmath,amssymb,amsthm,amscd,graphicx,mathtools}
\usepackage{hyperref}
\usepackage[subrefformat=parens,labelformat=parens]{subcaption}
\usepackage{suffix}
\usepackage{simpler-wick}
\usepackage[compat=1.1.0]{tikz-feynman}
\usepackage{slashed}
\input epsf.sty

\graphicspath{{figures/}}

\theoremstyle{definition}

\newcommand{\CC}{{\cal C}}
\newcommand{\CE}{{\cal E}}
\newcommand{\CF}{{\cal F}}

\newcommand{\CH}{{\cal H}}
\newcommand{\CI}{{\cal I}}

\newcommand{\CL}{{\cal L}}

\newcommand{\CO}{{\cal O}}

\newcommand{\CT}{{\cal T}}

\def\IZ{{\mathbb Z}}

\newcommand{\tr}{{\rm Tr}}
\newcommand{\re}{{\rm e}}
\newcommand{\ri}{{\rm i}}
\newcommand{\rd}{{\rm d}}

\newcommand{\mZ}{\mathsf{Z}}

\newcommand{\bpsi}{\boldsymbol{\psi}}

\newcommand{\be}{\begin{equation}}
\newcommand{\ee}{\end{equation}}
\newcommand{\ba}{\begin{aligned}}
\newcommand{\ea}{\end{aligned}}
\newcommand{\ben}{\begin{eqnarray}\displaystyle}
\newcommand{\een}{\end{eqnarray}}

\newcommand{\sectiono}[1]{\section{#1}\setcounter{equation}{0}}

\newdimen\tableauside\tableauside=1.0ex
\newdimen\tableaurule\tableaurule=0.4pt
\newdimen\tableaustep
\def\phantomhrule#1{\hbox{\vbox to0pt{\hrule height\tableaurule width#1\vss}}}
\def\phantomvrule#1{\vbox{\hbox to0pt{\vrule width\tableaurule height#1\hss}}}
\def\sqr{\vbox{%
  \phantomhrule\tableaustep
  \hbox{\phantomvrule\tableaustep\kern\tableaustep\phantomvrule\tableaustep}%
  \hbox{\vbox{\phantomhrule\tableauside}\kern-\tableaurule}}}
\def\squares#1{\hbox{\count0=#1\noindent\loop\sqr
  \advance\count0 by-1 \ifnum\count0>0\repeat}}
\def\tableau#1{\vcenter{\offinterlineskip
  \tableaustep=\tableauside\advance\tableaustep by-\tableaurule
  \kern\normallineskip\hbox
    {\kern\normallineskip\vbox
      {\gettableau#1 0 }%
     \kern\normallineskip\kern\tableaurule}%
  \kern\normallineskip\kern\tableaurule}}
\def\gettableau#1{\ifnum#1=0\let\next=\null\else
\squares{#1}\let\next=\gettableau\fi\next}

\newcommand{\bea}{\begin{eqnarray}\displaystyle}
\newcommand{\eea}{\end{eqnarray}}

\let\originalleft\left
\let\originalright\right
\renewcommand{\left}{\mathopen{}\mathclose\bgroup\originalleft}
\renewcommand{\right}{\aftergroup\egroup\originalright}

\newcommand{\figref}[1]{Fig.~\protect\ref{#1}}
\title{\Huge{\boldmath Trans-series from condensates}}

\author{Marcos Mari\~no$^a$ and Ramon Miravitllas$^b$}

\affiliation{
   ${}^a$D\'epartement de Physique Th\'eorique et Section de Math\'ematiques\\
  Universit\'e de Gen\`eve, Gen\`eve, CH-1211 Switzerland
   \vskip 0.01cm
   ${}^b$HUN-REN Wigner Research Centre for Physics\\
  Konkoly-Thege Miklós u. 29-33, 1121 Budapest, Hungary}

\emailAdd{Marcos.Marino@unige.ch} 
\emailAdd{ramon.miravitllas.mas@wigner.hu}

\abstract{The Shifman--Vainshtein--Zakharov (SVZ) sum rules provide a method to obtain 
trans-series expansions in many quantum field theories, in which exponentially small corrections 
are calculated by combining the operator product expansion with the assumption of vacuum condensates. In some solvable models, exact expressions for trans-series can be obtained from non-perturbative results, and this makes it possible to test the SVZ method by 
comparing its predictions to  these exact trans-series. In this paper we perform such a precision test in the example of the fermion self-energy in the Gross--Neveu model. Its exact trans-series expansion can be extracted from the large $N$ solution, at the first non-trivial order in $1/N$. It is given by an 
infinite series of exponentially small corrections involving factorially divergent power series in the 't Hooft parameter. We show that the 
first two corrections are associated to two-quark and four-quark condensates, and we reproduce the corresponding power series 
exactly, and at all loops, by using the SVZ method. In addition, the numerical values of the condensates can be 
extracted from the exact result, up to order $1/N$.}

\begin{document}
\maketitle
\flushbottom

\sectiono{Introduction}

In quantum field theory (QFT), perturbative series give the asymptotic expansion of observables at small coupling. 
There are many indications that this expansion can be upgraded to a 
{\it trans-series}, i.e. a generalization of the perturbative series which 
includes exponentially small corrections, in such a way that the exact value of the observable 
can be obtained by an appropriate resummation.  

Examples in QFT where we know explicitly the detailed structure of the trans-series are scarce. 
In some two-dimensional asymptotically free theories, 
one can compute certain observables exactly, and then show that this exact expression can 
be re-expressed or ``decoded" as a resummed trans-series. A beautiful realization of this idea 
was achieved in \cite{beneke-braun}, building on previous work \cite{david2,david3}, in the 
case of the two-point function of the non-linear sigma model 
at next-to-leading order in the $1/N$ expansion\footnote{The $1/N$ expansion, 
when used appropriately, gives a power series in $1/N$ where each term is an 
exact, non-perturbative function of the renormalized 't Hooft coupling, and not just a 
formal power series thereof. Sometimes we will use the expression ``exact results 
at large $N$," or similar ones, to refer to this type of non-perturbative functions.}. 
More recently, the free energy of integrable models coupled to a conserved charge 
was decoded as a resummed trans-series, even at finite $N$ \cite{mmr-an,mmr-theta,abbh2,bbv2,schepers,bbv3}. 

In generic QFTs, in which no analytic answer is known for the observables, 
physicists have devised two ways of upgrading the perturbative series to a 
trans-series. The first one is to add instanton corrections, coming from 
non-trivial saddle-points of the path integral. Instanton calculus is plagued 
with problems and it is fair to say that it is of limited use, except in 
supersymmetric theories or in very simple models. The second method 
to obtain trans-series can be applied to correlation functions in very 
general QFTs. It combines Wilson's operator product expansion (OPE) 
with some assumptions on the vacuum structure of the theory\footnote{The fact that the method of OPE with condensates  
leads to trans-series, in the sense of the theory of resurgence, was pointed out some time ago in \cite{stingl}. This was also observed more recently in \cite{shifman-renormalons}.}. More precisely, 
this method assumes the existence of {\it condensates}, or non-zero vacuum 
expectation values (vevs), for the operators appearing in the OPE. It 
was used by Politzer in \cite{politzer} to calculate the quark propagator in 
QCD beyond perturbation theory, 
and then extended and systematized in the famous QCD sum rules of Shifman, 
Vainshtein and Zakharov (SVZ) in \cite{svz, svz2}

An obvious question is whether the method of OPE and vacuum 
condensates provides the correct trans-series representation of correlation 
functions. Although this might be obvious to 
most practitioners in the field, there are various reasons for a detailed inquiry. For example, it could be 
the case that the OPE provides only an approximate parametrization of 
non-perturbative corrections, rather than the 
real thing. There has also been some debate concerning which form of the 
OPE should be used to calculate trans-series.  
Most of the sum rule calculations done by physicists are based on a simplified or ``practical" 
version of the OPE, in which the Wilson coefficients are calculated perturbatively, 
while the vevs of the operators contain the non-perturbative information, but it has been 
suggested \cite{svz-ope,svz-pr} that one might need more complicated procedures. Another 
important motivation to revisit these questions are the explicit results on non-perturbative corrections 
at finite $N$ obtained recently in integrable models \cite{mmr-an,mmr-theta,abbh2,bbv2,schepers,bbv3}, and 
whether condensate techniques 
can provide a first principle derivation of these results.

It is a good idea to ask foundational questions in simpler, 
solvable models where they can be answered with precision. 
In the case of the SVZ method this has been done in various papers, starting in the 1980s 
\cite{david2,david3, svz-ope,svz-pr, beneke-braun, sss}. These works usually focused on the 
two-dimensional non-linear sigma model at large $N$, and they extracted trans-series 
from exact results at leading and next-to-leading order in the $1/N$ expansion, as we mentioned above. Evidence was given that 
these trans-series are in agreement with the structure of the OPE, and in \cite{svz-ope, svz-pr,sss} this was 
verified in some cases by explicit calculations. However, in challenging examples like the one studied in \cite{beneke-braun}, 
in which the power corrections contain infinite, non-trivial perturbative series in the 't Hooft parameter, it 
was assumed that the OPE would reproduce these series, rather than verified explicitly. 

The goal of this paper is to provide a direct comparison between an exact trans-series 
and a standard calculation of power corrections. 
The original example of the two-dimensional non-linear sigma model is not the simplest one to perform 
such a comparison, and we focus instead on a fermionic cousin, the Gross--Neveu (GN) 
model \cite{gross-neveu}, where trans-series are of comparable complexity. The self-energy 
of elementary fermions (or ``quarks") in this model can be calculated at the first non-trivial order in 
the $1/N$ expansion, as an exact function of the external momentum and the 
mass gap \cite{cr-dr}. By using the Mellin transform techniques of \cite{beneke-braun} one can 
obtain an explicit trans-series representation of this function, involving an infinite series 
of power corrections. Schematically, we have 
\be
\Sigma(p)= \slashed p  \Sigma_0 (\lambda)+ \Lambda \Sigma_1 (\lambda)+\slashed p {\Lambda^2 \over p^2} \Sigma_2 (\lambda) +\cdots,
\ee
where $\Lambda$ is the dynamically generated scale and $\lambda$ is the 't Hooft coupling. The series $\Sigma_0(\lambda)$ is 
the perturbative series, but each power correction involves a factorially divergent series $\Sigma_n(\lambda)$, $n=1,2, \cdots$. 
If the OPE picture is correct, 
one should be able to reproduce these series by doing perturbation theory in 
the background of the appropriate vacuum condensates. This is precisely what 
we verify with complete success, and at {\it all} loops, 
for the first two power corrections, which are associated to the two-quark condensate 
and to the four-quark condensate (these are the terms with $n=1,2$ in the equation above, respectively). 
Once this is done, the values of the condensates at next-to-leading order (NLO) 
in $1/N$, which are unknown parameters in the sum rules, can be extracted from the large $N$ result. 

Our calculation provides a precise and direct test of the SVZ method in the 
GN model, at the first non-trivial order in $1/N$ expansion, and it illustrates various conceptual and practical 
issues of the method. For example, it shows that the four-quark condensate is not 
ambiguous at leading order in the $1/N$ expansion, due to factorization, but it is indeed 
ambiguous at subleading order, as expected from the 
results of \cite{david2,david3}. Our calculation is done with the ``practical" version of the 
OPE, which leads to the correct result in this example. An amusing 
spin-off result of this work is a diagrammatic derivation of the beta function of the model, 
at next-to-leading order in the $1/N$, which 
seems to be simpler than the approach usually followed in the literature \cite{pmp,ab}. 

We should mention that the paper 
\cite{landau} considered the chiral GN model, and 
compared the exact $1/N$ result for the propagator of the sigma particle 
to an OPE calculation with condensates, to leading order in the 't Hooft coupling. The resulting 
trans-series is much simpler than the ones considered here. We derive this result for the sigma propagator, in the GN model, 
at the end of section \ref{sec_4quark}. 

This paper is structured as follows. In section \ref{sec-gnbasics} we review or derive 
various basic results for the GN model which will be needed for the paper. In section \ref{sec-ts-1N} 
we obtain the trans-series expression for the exact two-point function at the first non-trivial order in the $1/N$ expansion. In section \ref{sec-condensate} we calculate the two-point function in perturbation theory with condensates, where we 
include the first two power corrections, and we reproduce 
exactly the trans-series derived from the $1/N$ expansion. Finally, in \ref{sec-conclusions} we 
conclude and present some questions and open problems. The Appendix collects some 
diagrammatic tools. It summarizes an important technique to calculate all-loop results in the $1/N$ expansion, due to Palanques-Mestre and Pascual \cite{pmp,pm}, which is used throughout the paper.

\sectiono{The Gross--Neveu model}
 \label{sec-gnbasics}
  
 The GN model is a two dimensional QFT, involving an $N$-uple of Dirac fermions with a quartic interaction, 
 which was introduced in \cite{gross-neveu} as a toy model for various important 
 physical phenomena. First of all, the GN model is asymptotically free\footnote{The first example of an asymptotically free 
 theory is in fact a close cousin of the GN model which was studied by Anselm in \cite{anselm}, 
 see \cite{shifman-anselm} for a historical appraisal.}. It can be solved in the large $N$ limit, where it can be shown that quantum effects lead to spontaneous symmetry breaking of a discrete $\IZ_2$ symmetry and the formation of a 
 bilinear fermion condensate.  For these reasons, 
the GN model can be seen as a toy model for the quark sector of QCD. In addition, the model is integrable at the quantum level,  its exact $S$-matrix has been conjectured in \cite{zz}, and its spectrum is extremely rich. 
In this section we will review some aspects of the model which we will need in our precision test of the 
SVZ method.

We will work in Minkowski space, and our choice of Dirac algebra in two dimensions is: 
\be
\gamma^0= \sigma^2, \qquad \gamma^1= \ri \sigma^1, \qquad \gamma^5= \sigma^3. 
\ee
The Lagrangian density describing the theory is 
\be
\CL= \ri  \overline{\bpsi} \cdot \slashed{\partial} \bpsi+ {g_0\over 2} \left(\overline{\bpsi} \cdot \bpsi  \right)^2, 
\ee
where $\boldsymbol{\psi}=(\psi_1, \cdots, \psi_N)$ is an $N$-uple of Dirac fermions. The model has a continuous $U(N)$ global symmetry, and a $\IZ_2$ discrete symmetry 
\be
\label{chirals}
\bpsi \rightarrow \gamma_5  \bpsi. 
\ee
In order to keep track of large $N$ counting, it is extremely useful to introduce an auxiliary scalar field $\sigma$ and write the GN Lagrangian as
\be
\label{sigma-lag}
\CL_\sigma= \ri  \overline{\boldsymbol{\psi}} \cdot \slashed{\partial} \boldsymbol{\psi}-{1\over 2 } \sigma^2 +{\sqrt{g_0}}  \sigma \overline{\boldsymbol{\psi}} \cdot \boldsymbol{\psi}. 
\ee
The original Lagrangian is obtained by integrating out $\sigma$. The symmetry (\ref{chirals}) reads now 
\be
\label{chiral-sigma}
\bpsi \rightarrow \gamma_5  \bpsi, \qquad \sigma \rightarrow -\sigma. 
\ee
One can also add a bare mass term for the fermions of the form 
\be
\label{fermion-mass}
\CL_{m_f} =- m_{0f} \overline \bpsi \bpsi, 
\ee
although we will consider the massless theory with $m_{0f}=0$ (as is well-known \cite{gross-neveu}, 
even when $m_{0f}=0$ a dynamical mass 
is generated at the quantum level and can be calculated in the $1/N$ expansion). 
From now one we will work with the Lagrangian (\ref{sigma-lag}). To write down the Feynman diagrams, we represent 
fermions by continuous lines and sigma particles by dashed lines. The fermion propagator in momentum space is given by 
\be
S_0(p)_{ij}^{\alpha \beta}= \left( {\ri \over \slashed p-m_{0f}} \right)^{\alpha \beta} \delta_{ij}. 
\ee
(Latin sub-indices are $U(N)$ indices, while Greek super-indices are spinor indices.)
The propagator of the $\sigma$ field is $-\ri$, and there is a single interaction vertex $\ri {\sqrt{g_0}}$.

The GN model is renormalizable and asymptotically free. We will almost always adopt the $\overline{\text{MS}}$ scheme and mostly work with bare fields, which we will simply denote by $\bpsi$. Renormalized fields will be denoted by $\bpsi_R$. The renormalization constants are defined as usual by
\be
\label{ren-constants}
\bpsi= Z_\psi^{1/2} \bpsi_R, \qquad g_0= (\nu^2)^{\epsilon/2} Z_g g, \qquad m_{0f}= Z_{m} m_f.
\ee
Our convention for $\epsilon$ is 
\be
\label{eps-conv}
d=2-\epsilon 
\ee
and
\be
\nu^2 = \mu^2 \re^{\gamma_E - \log(4\pi)},
\ee
where $d$ is the number of space-time dimensions in dimensional regularization and $\mu$ is the scale parameter.
The beta function is defined as 
\be
\beta(g; \epsilon)= -{\epsilon g \over 1+ g\partial_g \log Z_g}= -\epsilon g + \beta(g),
\label{beta-function}
\ee
with 
\be
\label{beta-ex}
\beta(g)= -\sum_{k \ge 0} \beta_k g^{k+2}, 
\ee
while 
\be
\gamma(g)= \beta(g;\epsilon) {\partial \log Z_\psi \over \partial g}, \qquad 
\gamma_m(g)= \beta(g;\epsilon) {\partial \log Z_m \over \partial g}
\ee
are the anomalous dimension of the field and the mass, respectively. The renormalization functions are 
known to four loops in conventional 
perturbation theory, see \cite{gracey-four} 
for recent results and references to the literature. However, we will work in the $1/N$ expansion. We define the 't Hooft parameter 
\be
\label{thooft-par}
\lambda= {g N \over \pi}
\ee
whose beta function is 
\be
\label{betal}
\beta_\lambda (\lambda; \epsilon)= {N \over \pi} \beta(g; \epsilon)=-\epsilon \lambda+\beta_\lambda (\lambda). 
\ee
The function $\beta_\lambda (\lambda)$ has a $1/N$ expansion at fixed 't Hooft coupling given by
\be
 \beta_\lambda (\lambda)=\sum_{j \ge 0}  \beta^{(j)}_\lambda (\lambda)  N^{-j} 
\ee
and similarly for  $\beta_\lambda (\lambda ;\epsilon)$. We note that 
\be
\label{leading-beta}
\beta_\lambda^{(0)}(\lambda;\epsilon)= -\epsilon \lambda- \lambda^2, \qquad \beta_\lambda^{(j)}(\lambda;\epsilon)=\beta_\lambda^{(j)}(\lambda), \quad j \ge 1. 
\ee
The first correction $\beta^{(1)}_\lambda (\lambda)$ is known in closed form \cite{gracey-eta} and is given by
\be
\label{beta1}
\beta_\lambda^{(1)}(\lambda)= \lambda^2 \left(1+ \int_0^\lambda {\Gamma(2+u) \over (2+u) \Gamma^3\left(1+{u \over 2} \right) 
\Gamma\left( 1-{u \over2} \right)} \rd u \right). 
\ee
We have similar results for the anomalous dimensions. 
The mass anomalous dimension has a $1/N$ expansion of the form
\be
\gamma_m(\lambda)= \sum_{j \ge 0} \gamma_m ^{(j)} (\lambda)N^{-j},   
\ee
where 
\be
\label{gamma0}
\gamma_m ^{(0)} (\lambda)=\lambda,
\ee
and the first non-trivial correction is given by \cite{gracey-mass}
\be
\label{gammafirst}
\gamma_m ^{(1)}(\lambda)=\chi(\lambda) - {\beta_\lambda^{(1)}(\lambda) \over \lambda}
   \ee
   where
   \be
   \label{chiF}
 \chi(\lambda)={\lambda  \Gamma (2+\lambda)\over (2+\lambda) \Gamma^3\left(1+{\lambda \over 2} \right) 
\Gamma\left( 1-{\lambda\over2} \right)}.
   \ee
  Finally, the field anomalous dimension has the $1/N$ expansion 
 \be
 \gamma(\lambda)=\sum_{j \ge 1}   \gamma^{(j)}(\lambda)N^{-j}, 
 \ee
where
\be
\label{ad-field}
\gamma^{(1)}(\lambda)= {\lambda^2\over 2} {1\over 2+ \lambda} {\Gamma(1+\lambda) \over \Gamma^3\left(1+{\lambda \over 2} \right) 
\Gamma\left( 1-{\lambda\over2} \right)}. 
\ee
As we will see, the functions (\ref{ad-field}), (\ref{chiF}) and (\ref{beta1}) will be obtained as spin-offs of our trans-series calculation for the perturbative, the two-quark condensate, and the four-quark condensate sectors, respectively. The renormalization constants in (\ref{ren-constants}) can be recovered from the renormalization functions. We have, for the coupling constant, 
\be
\label{zg-ex}
Z_g= \exp \left[ -\int_0^g {\rd u \over u} {\beta(u)\over \beta (u; \epsilon)} \right], 
\ee
while, for the field and mass renormalization, one finds 
\be
\label{zz-ex}
Z_\psi= \exp \left[ \int_0^g \rd u {\gamma (u)\over \beta (u; \epsilon)} \right], \qquad 
Z_m= \exp \left[ \int_0^g \rd u {\gamma_m (u)\over \beta (u; \epsilon)} \right].
\ee
The renormalization constants can also be obtained in a $1/N$ expansion by simply re-expressing everything in terms of the 't Hooft coupling. In particular, this coupling renormalizes as 
\be
\label{thooft-ren}
\lambda_0 = (\nu^2)^{\epsilon/2} Z_\lambda \lambda, 
\ee
where $Z_\lambda$ is the renormalization constant $Z_g$ expressed in terms of $\lambda$ and organized in a $1/N$ expansion. It is given explicitly as 
\be
\label{zlambda}
Z_\lambda= \exp \left[ -\int_0^\lambda {\rd u \over u} {\beta_\lambda(u)\over \beta_\lambda (u; \epsilon)} \right]. 
\ee

We will also need to renormalize the composite operators appearing in the OPE. We will denote 
renormalized composite operators by a bracket, $[\CO]$. The renormalization constants are defined by
\be
\label{op-mixing}
\CO_i= \mZ_{ij} \left[ \CO_j \right], 
\ee
where repeated indices are summed, and we have assumed mixing between a set of operators $[\CO_i]$, $i=1, \cdots, n$. Our convention for the matrix of anomalous dimensions is 
\be
\label{ad-matrix}
\boldsymbol{\gamma}= -\beta(g; \epsilon) {\partial  \mZ^{-1} \over \partial g} \mZ. 
\ee
We will consider the operator of dimension $1$
\be
\label{mass-term}
\overline \bpsi(x) \bpsi(x)
\ee
and the operators of dimension $2$, 
\be
\label{lag-ops}
K = \ri \overline {\bpsi}(x)  \cdot \slashed{\partial} \bpsi (x), \qquad V = g_0 \left(\overline{\bpsi}(x) \bpsi (x) \right)^2,
\ee
which appear in the Lagrangian. The renormalization of the fermion bilinear is straightforward, since it is the mass term 
appearing in the Lagrangian, and we have 
 \be
 \label{bil-ren}
\overline \bpsi(x) \bpsi(x) = Z_{\overline \bpsi \bpsi}  [\overline \bpsi(x) \bpsi(x)]
\ee
where (see e.g. \cite{collins})  
 \be
 \label{Zpp}
  Z_{\overline \bpsi \bpsi}= Z^{-1}_m. 
 \ee

 Let us now consider the operators $K$ and $V$. They mix under renormalization, and we can calculate the matrix $\mZ_{ij}$ very easily by following the method of \cite{gran, robertson}. In this method one starts with the bare and renormalized effective actions 
\be
\ba
\Gamma^0& = \int \rd^d x \left\{ A \ri  \overline{\bpsi} \cdot \slashed{\partial} \bpsi + B {g_0\over 2} \left(\overline{\bpsi} \cdot \bpsi \right)^2  + \cdots \right\}, \\
\Gamma& = \int \rd^d x \left\{ A Z_\psi \ri  \overline{\bpsi}_R \cdot \slashed{\partial} \bpsi_R + B Z_g Z_\psi^2 {g\over 2} \left(\overline{\bpsi}_R \cdot \bpsi_R \right)^2+ \cdots \right\}.  
\ea
\ee
The renormalization constants $Z_\psi$ and $Z_g$ are chosen so that divergences are reabsorbed, and we will fix them in such a way that 
\be
AZ_\psi= B Z_g Z_\psi^{2}=1. 
\ee
Let us consider the generating functional of 1PI Green functions with 
insertions of the bare operators $K$, $V$ at zero momentum. It can be obtained by acting with 
appropriate (functional) derivatives with respect to bare quantities on the bare effective action $\Gamma^0$:
\be
\ba
\Gamma_K^0 &=\left[ {1\over 2} \int \rd^d x \left( \psi_{i}^{\alpha} {\delta \over \delta \psi_{i}^{\alpha} (x)}+  \psi^{\dagger \, \alpha }_{i} {\delta \over \delta \psi^{\dagger \,\alpha}_{i} (x)} \right)-2 g_0 {\partial \over \partial g_0} \right] \Gamma^0, \\
\Gamma_V^0&= 2 g_0 {\partial \over \partial g_0}  \Gamma^0.  
\ea
\ee
The renormalized generating functionals for operator insertions of $[K]$ and $[V]$ can 
be similarly obtained by taking derivatives of $\Gamma$ with respect to renormalized quantities. The renormalization matrix $\mZ$ for the Lagrangian operators satisfies
\be
\begin{pmatrix} \Gamma^0_K \\ \Gamma^0_V \end{pmatrix} = \mZ \begin{pmatrix} \Gamma_K \\ \Gamma_V \end{pmatrix},
\ee
and a simple calculation shows that 
\be
\mZ= \begin{pmatrix} 1- 2 A' & - B' \\ 2 A' & 1+ B' \end{pmatrix},
\ee
where we have denoted 
\be
A'= g_0 {\partial \log A \over \partial g_0} 
\ee
and similarly for $B$. Explicit expressions for these quantities can be obtained from (\ref{zg-ex}), (\ref{zz-ex}), and one 
finds
\be
A'= {\gamma(g) \over \epsilon}, \qquad B'= {1\over \epsilon} \left( 2 \gamma(g)- {\beta(g) \over g} \right). 
\ee
We conclude that 
\be
\label{Zmatrix}
\begin{pmatrix} K \\ V \end{pmatrix}= \begin{pmatrix}  1- {2 \gamma \over \epsilon} & {1 \over \epsilon} \left( {\beta \over g}- 2 \gamma \right) \\[3mm]
{2 \gamma \over \epsilon} & 1+ {1 \over \epsilon} \left(2 \gamma- {\beta \over g} \right) \end{pmatrix} 
\begin{pmatrix} [K] \\ [V] \end{pmatrix}, 
\ee
and the matrix of anomalous dimensions is given by
\be
\label{gam-matrix}
\boldsymbol{\gamma}=  \left(
\begin{array}{cc}
 2 g \gamma '(g) &  \quad 2 g \gamma '(g)- \left( \beta '(g)-\frac{ \beta (g)}{g} \right) \\[3mm]
 -2 g \gamma '(g) &\quad  -2 g \gamma '(g) +\beta '(g)-\frac{\beta (g)}{g}
\end{array}
\right). 
\ee
One can verify explicitly from (\ref{Zmatrix}) that the operator $K +  V$ does not renormalize. This is a consequence of the fact that $K+V$ can be written as the product of an operator times an equation of motion \cite{collins}. Indeed, we have
 \be
 \label{EOM}
 \overline \bpsi {\delta S \over \delta \overline \bpsi}= K +  V,
 \ee
where $S$ is the action.
 
 \sectiono{Trans-series from the $1/N$ expansion}
\label{sec-ts-1N}

\subsection{An exact result for the self-energy} 

The GN model can be solved exactly at large $N$, and this means that one can 
calculate correlation functions as a systematic expansion in $1/N$ (see \cite{bkkw-un,bkkw} for an excellent 
presentation). In the large $N$ formulation, one 
integrates out the fermions in the action (\ref{sigma-lag}) and writes down the following effective action for the $\sigma$ field:
\be
\label{seff}
S_{\rm eff}= - \int \rd^2 x {\sigma ^2  \over 2 g_0} -\ri  N \tr \log (\ri S_0^{-1}), 
\ee
where we have rescaled $\sigma \rightarrow \sigma/{\sqrt{g_0}}$ as compared to (\ref{sigma-lag}), and 
\be
\label{S0}
S_0(\sigma)= \ri \left(\ri \slashed \partial - \sigma\right)^{-1}
\ee
is the free propagator for a Dirac fermion. This action has two saddle points at large $N$ in which $\sigma$ takes a constant value $\sigma_c=\pm m_0$, and the classical $\IZ_2$ symmetry (\ref{chiral-sigma}) is dynamically broken. %
The value of $m_0$ is determined by the gap equation 
\be
\label{gap}
{1\over N g_0} = {1\over m_0} \int {\rd^d k \over (2 \pi)^d} \tr \left[ {\ri \over \slashed k - m_0} \right].
\ee

The propagator for the fluctuations of the $\sigma$ field is defined as 
\be
\Delta^{-1}(x, y)= -{\ri \over N} {\delta^2 S_{\rm eff} \over \delta \sigma(x) \delta \sigma(y)}, 
\ee
evaluated at the large $N$ saddle point $\sigma_c=m_0$. In momentum space it is given by 
\be
\label{D-ex}
\Delta^{-1}(p;m_0)= {\ri \over 2 \pi} \xi \log\left[ {\xi+1 \over \xi-1} \right], 
\ee
where 
 \be
 \label{xi-def}
 \xi = {\sqrt{1- {4 m_0^2 \over p^2}}}.
 \ee
See Appendix \ref{pol-app} for some ingredients in the derivation of this formula. 
The large $N$ theory describes $\sigma$ particles interacting with fermions, 
the coupling is of order $N^{-1/2}$, and correlation functions 
can be computed in terms of large $N$ Feynman diagrams. The fermion self-energy has the 
following form:
\be
\label{sigma-self}
\Sigma(p)=m_0+ {1\over N} \left( \slashed p \Sigma_p + m_0 \Sigma_m \right), 
\ee
\begin{figure}
\centering
\includegraphics{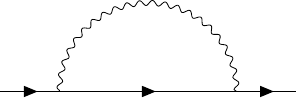}
\qquad \qquad
\includegraphics{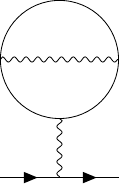}
\caption{Fermion self-energy diagrams at large $N$. The wavy line corresponds to the propagator of the $\sigma$ field, defined in \eqref{D-ex}.}
\label{exact-selfenergy-diagrams}
\end{figure}%
where $\Sigma_{p,m}$ have a $1/N$ expansion 
\be
\label{sigma-fexp} 
\Sigma_{p,m}= \sum_{j \ge 1} \Sigma_{p,m}^j N^{-j+1}. 
\ee
The leading order terms can be computed from the diagrams in \figref{exact-selfenergy-diagrams}, and they are given by \cite{bkkw-un,cr-dr,fnw2}
\be
\label{sigma-ss}
\ba
\Sigma^1_p&= {1\over p^2} \int {\rd^2  k \over (2 \pi)^2} {p^2+ k\cdot p \over (p+k)^2-m_0^2} \Delta(k;m_0), \\
\Sigma^1_m&= \int {\rd^2  k \over (2 \pi)^2} \left({\Delta(k;m_0) \over (p+k)^2-m_0^2} - {2 \pi \ri \over k^2- 4 m_0^2} \right). 
\ea
\ee
These integrals are divergent and they have to be regularized and renormalized. In \cite{cr-dr}, Campostrini and Rossi 
found explicit, finite expressions for them by using a sharp momentum cutoff (SM) 
regularization scheme. In the SM scheme, one first performs a Wick rotation to Euclidean space and computes the angular integral. Then, the resulting integrand is Taylor expanded at infinity. The terms which lead to a divergence are simply subtracted, but in order to avoid IR divergences in the 
subtracted pieces one has to introduce an IR cutoff $M$ which plays the role of the renormalization scale $\mu$ in dimensional regularization, see \cite{cr-review,biscari} for more details. The renormalized self-energy in the SM scheme can be written in terms of the functions
\be 
\ba
A(x)&= {1\over 4 x} \int_0^\infty \left( \xi_y \log\left[ {\xi_y+1 \over \xi_y-1} \right] \right)^{-1} \left(1- {1+y-x \over {\sqrt{ (x+y+1)^2 - 4 xy }}} \right)\, \rd y, \\
B(x)&=  {1\over 2} \int_0^\infty \left( \xi_y\log\left[ {\xi_y+1 \over \xi_y-1} \right] \right)^{-1} \left( {1 \over {\sqrt{ (x+y+1)^2 - 4 xy }}} +{1 - \xi_y \over 2}\right)\, \rd y,
\ea
\ee
where
\be
\xi_y= {\sqrt{1 +{4 \over y}}}. 
\ee
Then the renormalized self-energy has the form,  
\be
\label{smse}
\Sigma^{\rm SM}(p)=m+ {1\over N} \left( \slashed p \Sigma_p^{\rm SM} + m \Sigma_m^{\rm SM}\right), 
\ee
where $\Sigma^{\rm SM}_{p,m}$ are given, at leading order in the $1/N$ expansion,  
\be
\label{leadSM}
\Sigma^{{\rm SM}}_p = - A\left( -{p^2 \over m^2} \right)+ \CO( N^{-1}), \qquad \Sigma^{{\rm SM}}_m =-B \left( -{p^2 \over m^2} \right)+ \CO( N^{-1}). 
\ee
In (\ref{smse}), $m$ is the mass gap, which differs from $m_0$ in $1/N$ corrections: 
\be
m= m_0 + {m_1 \over N}+ \CO\bigl(N^{-2}\bigr). 
\ee
$m_1$ can be calculated in terms of the 't Hooft coupling in the SM scheme, see \cite{cr-dr} for details. It is also possible to calculate the beta function and the anomalous dimension of the field in the SM scheme. They are given by \cite{cr-dr}
\be
\label{sm-scheme}
\beta_{\rm SM}(\lambda)= -\lambda^2 + {1\over 2N} (2+\lambda) \lambda^2+ \CO\bigl(N^{-2}\bigr), \qquad 
\gamma_{\rm SM}(\lambda)= \CO\bigl(N^{-2}\bigr). 
\ee
The result for the anomalous dimension follows immediately from the fact that $\Sigma^1_p$ in (\ref{sigma-ss}) is finite.

The SM scheme is the most useful one to perform calculations in the exact $1/N$ expansion, but in conventional perturbation theory we 
will use the $\overline{\text{MS}}$ scheme. We therefore have to compare quantities computed in different schemes. Let us introduce the dynamically generated scale in the $\overline{\text{MS}}$ scheme, $\Lambda$, in the conventions of \cite{fnw1,fnw2}. It is given by 
\begin{equation}
\Lambda=\mu \left( {\beta_0 g \over 2} \right)^{-\beta_1/\beta_0^2} \re^{-1/( \beta_0 g)} 
\exp\left( -\int_0^g \left\{ {1\over \beta(x)}+ {1\over \beta_0 x^2} -{\beta_1 \over \beta_0^2 x} \right\} \rd x  \right),
\end{equation}
where $\beta_0$, $\beta_1$ are the first coefficients of the beta function in (\ref{beta-ex}). By comparing calculations in the SM scheme and in the $\overline{\text{MS}}$ scheme, \cite{fnw1,fnw2} were able to relate the mass gap $m$ to $\Lambda$. One finds:
\be
\label{m-lambda}
m = \left( 1+ {1\over N} \left( \log 2 -{\gamma_E \over 2} +{1\over 2} \right) + \cdots \right) \Lambda. 
\ee
For the purposes of this paper, we want to compare the self-energy computed in different schemes. We first recall that, 
in a change of scheme characterized by a coupling $g$ to a new one characterized by $g'$, $\ell$-point Green functions pick a factor $\zeta^\ell (g)$, where $\zeta(g)=1 +\CO(g)$ is determined by the equation (see e.g. \cite{collins})
\be
\gamma'(g')- \gamma(g)= - 2 \beta(g) {\partial \over \partial g} \log \zeta(g), 
\ee
and $\gamma'(g')$ is the anomalous dimension in the new scheme. In our case, 
using the SM anomalous dimension in (\ref{sm-scheme}), we find that 
\be
\label{zeta-1N}
\zeta(\lambda)= 1+ {1\over 2 N} \int_0^\lambda {\gamma^{(1)}(u) \over \beta^{(0)}_\lambda (u)} \rd u+ \CO\bigl(N^{-2}\bigr).  
\ee
This result will be crucial for the comparison of the two calculations.

\subsection{Trans-series expression}

The functions $A(x)$, $B(x)$ give the exact, non-perturbative result for the 
self-energy of the GN model up to order $1/N$, in the SM scheme. We will 
now show that these functions can be written as Borel-resummed 
trans-series, by following the method developed in \cite{beneke-braun} for the self-energy 
of the non-linear sigma model (see also \cite{mmr-trans} for 
further details on the strategy and results of \cite{beneke-braun}). For background on 
trans-series, Borel resummation, and the theory of resurgence, we refer the reader to \cite{mmlargeN,mmbook,ss, abs}. 

Let us first focus on $A(x)$. The first step is to write
\begin{equation}
\left[\log\left( \frac{\xi_y+1}{\xi_y-1} \right) \right]^{-1} = \int_0^\infty \rd t \left[ \frac{\xi_y-1}{\xi_y+1} \right]^t.
\end{equation}
We define the following functions, given as Mellin transforms, 
\begin{equation}
K_i(s,t) = \int_0^\infty \xi_y^i \left[ \frac{\xi_y-1}{\xi_y+1} \right]^t y^{s-1} \rd y, \qquad i \in \IZ, 
\end{equation}
which can be computed as quotients of $\Gamma$ functions. We will use 
\be
K_{-1}(s,t)=\frac{\Gamma(2s+1)\Gamma(-s+t)}{\Gamma(s+t+1)}. 
\ee
We also define
\begin{equation}
\begin{aligned}
M(s) &= \int_0^\infty \frac{y^2}{\sqrt{(x+y+1)^2-4xy}} y^{s-1} \rd y\\
&= (1+x)^{s+1} B(s+2,-1-s) {}_2F_1\left( s+2, -1-s ; 1 ; \frac{x}{1+x} \right).
\end{aligned}
\label{eq_sqrt_mellin}
\end{equation}
Then, by using the properties of the Mellin transform and its inverse, we obtain the following representation for the function $A(x)$,
\begin{equation}
\begin{multlined}[b][0.75\textwidth]
A(x) = \frac{1}{4x} \int_0^\infty \rd t \Biggl[ - \int_{C_1} \frac{\rd s}{2\pi\ri} K_{-1}(-s,t)M(s)\\
+ (x-1) \int_{C_2} \frac{\rd s}{2\pi\ri} K_{-1}(-s,t) M(s-1) + K_{-1}(1,t) \Biggr]. 
\label{eq_selfenergy_A_mellin}
\end{multlined}
\end{equation}
In this equation, $C_1$ is the line $c+\ri\mathbb{R}$ with $c \in (-2,-1)$. For $C_2$, we must consider instead $c \in (-1,0)$\footnote{For $C_1$, the value of $c$ has to be in the region of convergence of the integral $M(s)$ appearing in the first line of \eqref{eq_sqrt_mellin}. For $C_2$, we have to consider instead $M(s-1)$, so the region of convergence gets displaced by 1. The values of $c$ will be crucial to determine which singularities have to be included when we deform the contours to the left of the complex plane.
}. 
We now expand the hypergeometric function in (\ref{eq_sqrt_mellin}) at large $x$: 
\begin{equation}
M(s) = x^{s+1} \sum_{k\ge 0} a_k(s) x^{-k},
\end{equation}
with coefficients
\begin{equation}
a_k(s) = \frac{(-1-s)_k \Gamma(s+2)}{k!^2 \Gamma(s+2-k)}\biggl[ \log(x) +2 \psi(k+1) - \psi(-s+k-1) - \psi(s+2-k) \biggr].
\end{equation}
We must also consider the combination
\begin{equation}
\begin{aligned}
-M(s) +(x-1)M(s-1) &= x^{s+1} \sum_{k\ge 0} \biggl( -a_k(s) + a_k(s-1) - a_{k-1}(s-1) \biggr)x^{-k}\\
&= x^{s+1} \sum_{k\ge 0} b_k(s) x^{-k},
\end{aligned}
\end{equation}
with the convention that $a_{-1}(s) = 0$. Explicitly, the coefficients $b_k(s)$ are given by
\begin{align}
b_0(s) &= 2/(1 + s),\\
b_k(s) &= \frac{2k\Gamma (s+1) \Gamma (k-s-1)}{k!^2 \Gamma (-s) \Gamma (s+1-k)} \biggl[\log(x) +2 \psi(k)+\frac{1}{k} -\psi(-s+k-1) -\psi(s-k+1)\biggr].
\label{eq_bk_coeff}
\end{align}
Eq.~\eqref{eq_selfenergy_A_mellin} is then written as
\begin{equation}
\begin{multlined}[b][0.75\textwidth]
A(x) = \frac{1}{4} \int_0^\infty \rd t \Biggl[ \sum_{k\ge 0} \int_{C_2} \frac{\rd s}{2\pi\ri} K_{-1}(-s,t)  b_k(s) x^{-k+s}\\
+ \int_{C_{-1}} \frac{\rd s}{2\pi\ri} K_{-1}(-s,t)  a_k(s) x^{-k+s} + \frac{K_{-1}(1,t)}{x} \Biggr],
\label{eq_selfenergy_A_mellin_2}
\end{multlined}
\end{equation}
where $C_{-1}$ is a small circle around $s=-1$. We compute the integral in $s$ by using the residue theorem. Deforming the contour $C_2$ to the left, we have contributions from:
\begin{enumerate}
\item Poles of $K_{-1}(-s,t)$ at $s = -t-j$,  $j\in\mathbb{Z}_{\ge 0}$. We group the poles contributing with a factor $(-x)^{-n} x^{-t}$, with fixed $n=k+j$ ($0\le k \le n$, $0 \le j \le n$). In this way, we define the functions
\begin{equation}
\CE_n(t) = \frac{(-1)^n}{4} \sum_{j=0}^n b_{n-j}(-t-j) \textrm{Res}_{s=-t-j} K_{-1}(-s,t).
\end{equation}
%

\item Poles of $b_k(s)$ at $s=-j-1$, $j\in\mathbb{Z}_{\ge 0}$, coming from the last digamma function in \eqref{eq_bk_coeff}.\footnote{Naively, there are more poles coming from the digamma function in $b_k(s)$, but they have residue 0. The case $k=0$ is an exception in which we only have a single pole at $s=-1$.} In this case, we group the poles contributing with a factor $-(-x)^{-n}$, with fixed $n=k+j+1$ ($0\le k \le n$, $0\le j \le n-1$). In this way, we define
\begin{multline}
\CH_n(t) = -\frac{(-1)^n}{4} \biggl[ K_{-1}(1,t)\textrm{Res}_{s=-1}\bigl(b_{n-1}(s)+a_{n-1}(s)\bigr)\\
+  \sum_{j=1}^{n-1} K_{-1}(j+1,t)\textrm{Res}_{s=-j-1}b_{n-j-1}(s) \biggr], \qquad n\neq 1,
\end{multline}
with the convention that $a_{-1}(s)=b_{-1}(s) = 0$. 

For $n=1$, we have to add the contribution from the term 
\be
\tfrac{1}{4}K_{-1}(1,t) = \dfrac{1}{2t(t^2-1)}
\ee
 in \eqref{eq_selfenergy_A_mellin_2}.
\end{enumerate}
With these conventions, we obtain the final result
\begin{equation}
A(x) = \sum_{n\ge 0} (-x)^{-n} \int_0^\infty \rd t  \bigl( x^{-t} \CE_n(t) - \CH_n(t) \bigr). 
\label{eq_A_transseries}
\end{equation}
We now recall that, in order to obtain the self-energy $\Sigma^{\rm SM}_p$, we have to set $x=-p^2/m^2$. 
We introduce the variable $\lambda$ as
\begin{equation}
-{m^2 \over p^2} = \re^{-2/\lambda}
\end{equation}
as well as the Borel variable $y = 2t$. Let us note the important fact that, at leading order in the $1/N$ expansion, 
$\lambda$ can be identified with the running 't Hooft parameter  
$\lambda(p)$ at the scale set by $p$ in the $\overline{\text{MS}}$ scheme, since at this order 
one has $m \approx \Lambda$. We also define the functions $E_n(y)$, $F_n(y)$, $G_n(y)$ and $H_n(y)$ as  
\be
\ba
E_n(y)&= \frac{\CE_n(y/2)}{2}= {1 \over \lambda}F_n(y) + G_n (y), \\
H_n(y)&= \frac{\CH_n(y/2)}{2}. 
\ea
\ee
We can then write
\begin{equation}
\label{Aint-alt}
A\left(-{p^2 \over m^2} \right) = \sum_{n\ge 0} \left( {m^2 \over p^2} \right)^n  \int_0^\infty \rd y  \left( \re^{-y/\lambda} E_n(y)- H_n(y) \right).
\end{equation}
One finds the explicit expressions, for $n=0,1$, 
\begin{equation}
\begin{aligned}
E_0(y)&= \frac{1}{2}\frac{1}{2-y}, &
H_0(y)&= 0,\\
E_1(y)&= {1\over \lambda}{y \over 4} -\frac{1}{y}-\frac{1}{2} + \frac{y}{8} \left[ 1 - 2\gamma_E - \psi\left(\frac{y}{2}\right) - \psi\left(-\frac{y}{2}\right) \right], &
H_1(y) &= \frac{4}{y(y-2)(y+2)},
\end{aligned}
\label{EnHn}
\end{equation}
where $\psi(x)$ is the digamma function. It is possible to check, on a case by case basis, the equalities
\begin{equation}
\textrm{Res}_{y=2k} G_n(y) = (-1)^k \textrm{Res}_{y=2k} H_{n+k}(y), \quad k,n\in\mathbb{Z}_{\ge 0},
\end{equation}
which are needed to have a complete cancellation of poles in the integral \eqref{Aint-alt}, after summing over all $n \ge 0$. However, the integrand in \eqref{Aint-alt} is singular for a fixed $n$. This allows us to write a formal trans-series out of the above expression, as follows. Let us define 
\be
r_{n,k} = {\rm Res}_{y=2k} \, H_n(y). 
\ee
The first step is to rearrange the integral for fixed $n$ as 
\be
\label{ex-borel}
\int_0^{\infty \re^{\ri\theta}} \left\{ \re^{-y/\lambda} \left( \frac{F_n(y)}{\lambda} + \widehat G_n(y) \right) - \left( H_n(y) - r_{n,0} {\re^{-y/\lambda} \over y} \right) \right\} \rd y,
\ee
where we have introduced the function 
\be
\widehat G_n(y)= G_n(y) - \frac{r_{n,0}}{y} 
\ee
which is regular at the origin. 
Note that for fixed $n$ the integrand has singularities for positive values of $y$, and we 
have deformed the integration contour slightly above or below 
the positive real axis with a small angle $\theta$, to make sense of the integral. The integrand of (\ref{ex-borel})
\be
F_n(y) + \lambda \widehat G_n(y)= \sum_{k \ge 0} {y^k \over k!} \left( F_n^{(k)}(0)+ \lambda \widehat G_n^{(k)}(0)  \right), 
\ee
can be regarded as the Borel transform of the factorially divergent series (our convention for the Borel transform is as in \cite{mmbook})
\be
\label{n-trans}
\varphi_n(\lambda)=\sum_{k \ge 0} \left(\lambda^k F_n^{(k)}(0)+ \lambda^{k+1} \widehat G_n^{(k)}(0) \right)=F_n(0)+ \sum_{k \ge 0} \left( F_n^{(k+1)}(0) + \widehat G_n^{(k)}(0)  \right) \lambda^{k+1}, 
 \ee
 and we can write
 \be
 \int_0^{\infty \re^{\ri\theta}} \re^{-y/\lambda} \left( \frac{F_n(y)}{\lambda} + G_n(y) - \frac{r_{n,0}}{y} \right)  \rd y = s_\pm \left( \varphi_n \right)(\lambda), 
  \ee
 where $s_\pm$ are lateral Borel resummations (see e.g. \cite{mmbook} for a definition of these). In this and similar 
 expressions in the following, the $\pm$ sign 
 is correlated with the sign of $\theta$ in the contour deformation. 
 
Let us now consider the second piece, involving $H_n(t)$. A simple calculation shows that
\be
\label{Hint}
-\int_0^{\infty \re^{\ri\theta}} \left( H_n(y) - r_{n,0} \frac{\re^{-y/\lambda}}{y} \right)\rd y =r_{n,0}\log(\lambda)+ c_n  \pm \ri\pi \sum_{k=1}^n r_{n,k}, 
\ee
where the constant $c_n$ is defined as the principal value integral
\begin{equation}
c_n = -{\rm{P}} \int_0^\infty \left( H_n(y) -r_{n,0} \frac{ \re^{-y}}{y} \right) \rd y.
\end{equation}
It is natural to include the logarithmic and constant pieces in the trans-series, 
so that the total, factorially divergent series for each $n$ is given by
\be
\Phi_{2n}(\lambda)=r_{n,0}\log(\lambda)+ c_n  \pm \ri\pi \sum_{k=1}^n r_{n,k}+ \sum_{k \ge 0} \left(\lambda^k F_n^{(k)}(0)+ \lambda^{k+1} \widehat G_n^{(k)}(0) \right), \qquad n \ge 0. 
\ee
We have labelled these formal series in $\lambda$ 
with an even index, $2n$, since they multiply even powers of $m$, $m^{2n}$, in the 
trans-series, and as we will see the function $m B(x)$ leads to odd powers $m^{2n+1}$. 
For $n=0,1$ we use the expressions in \eqref{EnHn} to obtain
\be
\label{ts1}
\ba
\Phi_0(\lambda)&=\sum_{k \ge 1} {(k-1)! \over 2^{k+1}} \lambda^{k},\\
\Phi_2(\lambda) &=  \gamma_E + \log(2) \pm {\ri \pi \over 2} - \log(\lambda) -{\lambda \over 4}+ {\lambda^2 \over 8} +\sum_{k \ge 1} {(2k+1)! \over 2^{2k+2}}  \zeta(2k+1)\lambda^{2k+2}. 
\ea
\ee
The series $\Phi_0(\lambda)$ corresponds to the perturbative sector of the self-energy. It has appeared already in \cite{cr-dr}, since it gives the classical 
asymptotic expansion of the function $A(x)$. On the other hand, $\Phi_2(\lambda)$ is the first trans-asymptotic correction to $A(x)$ and multiplies a power 
correction of order $m^2/p^2$. In addition, it is ambiguous. The ambiguity has a very simple interpretation, in view of the above analysis. The Borel transform of the perturbative series $\Phi_0(\lambda)$ has a singularity at $y=2$, therefore its two lateral Borel resummations are different. However, the choice of the constant term in the trans-series $\Phi_2(\lambda)$ can be correlated with the choice of lateral resummation of $\Phi_0(\lambda)$, so that the final result is the same for both choices. This is an illustration 
of the general phenomenon noted by David in \cite{david2}. 

So far we have analyzed the trans-series expression for $A(x)$. In the case of $B(x)$, the trans-series structure can be obtained 
from the observation that \cite{cr-dr}
\be
B(x)= {1\over 2 (1+ 3/x)} \left( 4 A(x) +{S(x) \over x} \right), 
\ee
where 
\be
S(x)=  \int_0^\infty \rd y \left(  \log\left[ {\xi_y+ 1  \over \xi_y-1}  \right]\right)^{-1} \left[  {y  \xi_y \over {\sqrt{(x+y+1)^2-4 x y}}} -1+
  {x+ 1 \over 2} \left( {1\over \xi_y}-1\right) \right].  
  \ee
The trans-series structure of $S(x)/x$ is known from \cite{beneke-braun,mmr-trans}, and it 
involves corrections in even powers of $m$. Therefore the 
trans-series structure for $B(x)$ follows from the results for $A(x)$ and $S(x)/x$. For the purposes of this paper, 
only the classical asymptotic series of $B(x)$ will be needed. It is given by the formal series in the 't Hooft parameter
\be
\label{ts2}
\Phi_{1}(\lambda)=  {1\over 2} -{\gamma_E \over 2} -{1\over 2} \log(2) + {1\over 2} \log(\lambda)+ \sum_{k \ge 1} {(2k)! \over 2^{2k+1}} \zeta(2k+1) \lambda^{2k+1}. 
\ee

Let us summarize the results in this section. The fermion self-energy $\Sigma(p)$ in the 
SM scheme, at order $1/N$, is given by an exact expression which is a function of the external 
momentum $p$ and the mass gap $m$. By using the Mellin transform techniques of \cite{beneke-braun}, 
this expression can be decoded as a trans-series. This involves a perturbative part, given by the series 
$\Phi_0(\lambda)$, a power correction proportional to $m \sim \Lambda$ and involving the series 
$\Phi_{1}(\lambda)$ in (\ref{ts2}), and a power correction proportional to $m^2 \sim \Lambda^2$ 
and involving the series $\Phi_{2}(\lambda)$ in (\ref{ts1}). More precisely, we have
\be
\label{ts-sigma}
\Sigma^{\rm SM} (p) \sim  - {1\over N} \slashed p \Phi_0(\lambda) + m \left( 1- {1\over N}  \Phi_{1}(\lambda)  \right)
 -{1\over N} \slashed p {m^2 \over p^2} \Phi_2(\lambda)+ \cdots.
\ee
We note that the l.h.s. of this equation is a well-defined function of $p$, while on the r.h.s. we 
have a trans-series representation. The first term is the perturbative 
series, while the second and the third term are non-perturbative power corrections. The dots refer to higher 
order power corrections, and to higher order corrections in $1/N$. Thanks to the exact large $N$ analysis, we have precise, 
all-loop predictions for the power series attached to each of these power corrections.

We should be able to 
reproduce the first term in the r.h.s. of (\ref{ts-sigma}) 
from a conventional perturbative calculation in, say, the ${\overline{\text{MS}}}$ 
scheme. This was essentially verified in \cite{cr-dr}, although we will present a more detailed 
calculation in the next section. In addition, if the method of OPE with vacuum condensates 
proposed in \cite{politzer,svz, svz2} gives the correct trans-series representation of observables, 
the power corrections in the r.h.s. of (\ref{ts-sigma}) should be 
calculable with the SVZ approach, up to unknown overall constants related to the values of the 
vacuum condensates. We will also verify this in the next section.

\sectiono{Trans-series from condensates}
\label{sec-condensate}
In this section we will calculate the series $\Phi_{0, 1,2}(\lambda)$ in perturbation theory with condensates. We will always work in the ${\overline{\text{MS}}}$ scheme. 

\subsection{Perturbative series}
\begin{figure}[!ht]
\centering
\includegraphics[width=10cm,trim={0 1.6cm 0 2.2cm},clip]{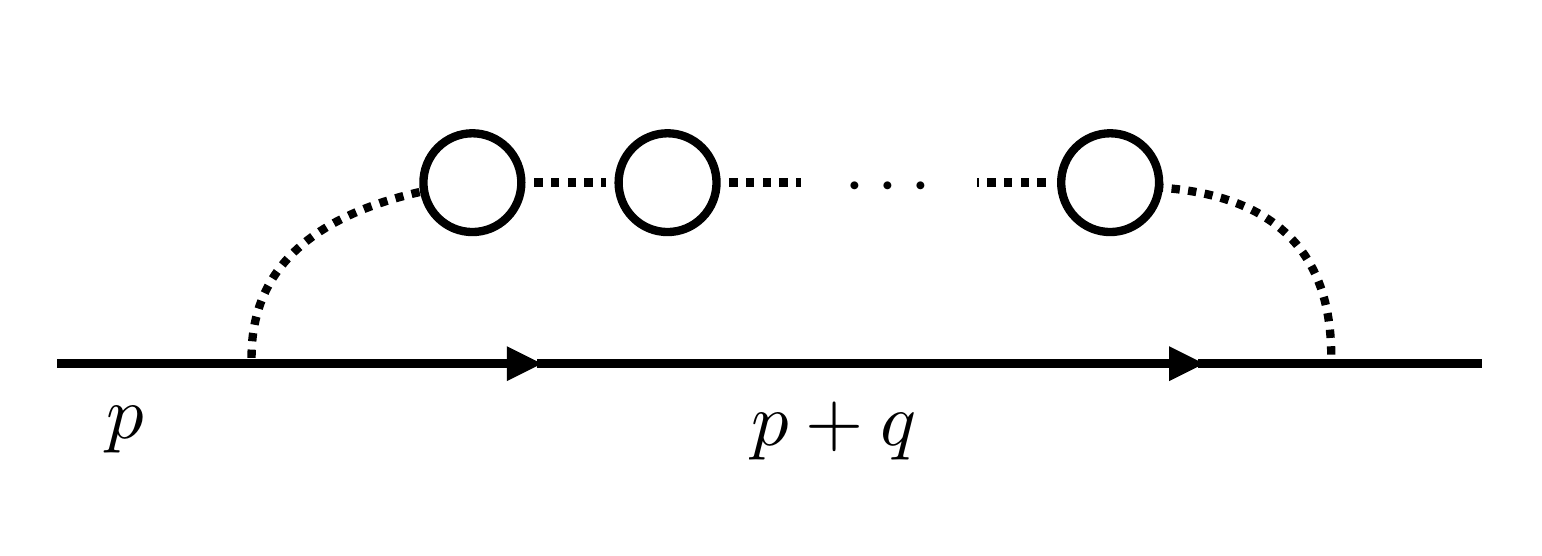}
\caption{The $1/N$ correction to the fermion self-energy is given by a chain of $n$ bubbles.}
\label{chain-bubble-fig}
\end{figure}
\noindent
The first step is to compute the leading $1/N$ correction to the fermion 
self-energy, at all orders in the 't Hooft coupling constant. This is 
a standard renormalon calculation, since the $n$-th order correction is due 
to a chain of $n$ fermion loops or ``bubbles," linked by $\sigma$ propagators, as 
shown in \figref{chain-bubble-fig}. Using the Feynman rules, one finds that the contribution of $n$ loops is 
\be
(\ri {\sqrt{g_0}})^{2n+2} (-\ri)^{n+1} N^n \Pi^n(q^2)= {\ri \over N} \left( \ri \Pi(q^2) \right)^n (\pi \lambda_0)^{n+1},
\label{chain-bubble}
\ee
where we have introduced the bare 't Hooft parameter $\lambda_0$ as in (\ref{thooft-par}), and $\Pi(q^2)$ is the fermion polarization loop (\ref{pi-massless}). We now express $\lambda_0$ in terms of the renormalized coupling constant through (\ref{thooft-ren}), 
and we sum the 
geometric series of bubbles to 
\be
{\ri \over N} {1\over (\pi \lambda (\nu^2)^{\epsilon/2})^{-1} Z_\lambda^{-1} - \ri \Pi (q^2)}. 
\ee
If we now take into account that 
\be
-\ri \Pi(q^2) = -{1\over \pi \epsilon} + \cdots,
\ee
we deduce that the renormalization constant $Z_\lambda$ is given, at large $N$, by 
\be
\label{zl}
Z^{-1}_\lambda= 1+ {\lambda \over \epsilon} + \cdots, 
\ee
in agreement with the result (\ref{leading-beta}) for the beta function at leading order in the $1/N$ expansion. 

We will write the (bare) self-energy as in (\ref{sigma-self}),  
\be
\Sigma(p)\sim {1\over N} \slashed p \Sigma_p, 
\ee
where the asymptotic sign $\sim$ emphasizes that our calculation will lead to a representation in terms of 
formal power series. 
If we define
\be
\label{i0-int}
\CI_0(n)= {1\over \slashed p} \pi^n  \int {\rd^d q \over (2 \pi)^d} {(\slashed p + \slashed q) \over (p+q)^2} ( \ri  \Pi(q^2))^{n-1},
\ee
it is easy to see that the leading term in the $1/N$ expansion of $\Sigma_p$ is 
\be
\label{sum-sigmap}
\Sigma_p^{{\rm P},1} = \sum_{n \ge 2} \ri \CI_0(n) \lambda_0^n,  
\ee
where we have added a superscript ${\rm P}$ to indicate that this is the perturbative result, and the superindex $1$ means that, as in 
(\ref{sigma-fexp}), this is the term of order $1/N$ in the $1/N$ expansion. We can perform the integral (\ref{i0-int}) explicitly by using (\ref{gen-loop}), and we find
\be
\CI_0(n) =  {\ri \over 4} 
\left(-\frac{p^2}{4\pi}\right)^{-n\epsilon/2} \frac{\Gamma \left(1-\frac{\epsilon }{2}\right) \Gamma \left(\frac{n \epsilon }{2}\right) \Gamma \left(1-\frac{n \epsilon }{2}\right)}{\Gamma \left(\frac{(n-1)\epsilon}{2} \right) \Gamma \left(2-\frac{(n+1)\epsilon}{2} \right)} \left[ -\frac{1}{2}  \frac{\Gamma\left(1-\frac{\epsilon}{2}\right)\Gamma\left(1+\frac{\epsilon}{2}\right)\Gamma\left(-\frac{\epsilon}{2}\right)}{\Gamma(1-\epsilon)} \right]^{n-1}.
\label{eq_coeff_self_energy}
\ee
We can now use the techniques of Appendix \ref{pmp-trick} to write the sum (\ref{sum-sigmap}) in terms of the structure function
\be
\label{fxy}
F(x,y) = -\frac{1}{2} \left(-\frac{p^2}{4 \pi \nu^2} \right)^{-y/2} \frac{\Gamma \left(1-\frac{x}{2}\right) \Gamma \left(1+\frac{y}{2}\right) \Gamma \left(1-\frac{y}{2}\right)}{\Gamma \left(\frac{y-x}{2}\right) \Gamma \left(2-\frac{y+x}{2}\right)} \left[\frac{\Gamma \left(1+\frac{x}{2}\right) \Gamma^2 \left(1-\frac{x}{2}\right)}{\Gamma (1-x)}\right]^{y/x-1}.
\end{equation}
This function contains both the divergent and the finite part of the bare self-energy. 
To renormalize the self-energy we need to introduce the renormalization of the field, which has the $1/N$ expansion
\be
Z_\psi= 1+ {1\over N} Z_\psi^{(1)}+ \cdots.
\ee
The renormalized self-energy is then given by  
\be
\label{self-p-ren}
\Sigma_{p,R}^{{\rm P},1} =\Sigma_{p}^{{\rm P},1} - Z_\psi^{(1)}. 
\ee
By using the results of Appendix \ref{pmp-trick} we find  
\be
\label{divzpsi}
Z_\psi^{(1)}=\left[\Sigma_{p}^{{\rm P},1} \right]_{\rm div} =\left[ F_0(\epsilon)\log\left(1+{\lambda \over \epsilon} \right) \right]_{\rm div}, 
\ee
where the function $F_0(\epsilon)=F(\epsilon, 0)$ is given by 
\be
\label{F0}
F_0(\epsilon)= {\epsilon \over 2} {1\over 2-\epsilon} {\Gamma(1-\epsilon) \over \Gamma^3\left(1-{\epsilon \over 2} \right) 
\Gamma\left( 1+{\epsilon \over2} \right)}
\ee
and we have taken into account that $F_{0,0}=F(0,0)=0$. 
We can now use (\ref{use-id}) to obtain the anomalous dimension of the field, at the first non-trivial order in $1/N$, as 
\be
\gamma^{(1)}(\lambda)=-\lambda F_0(-\lambda)
\label{anomalous_F0}
\ee
which reproduces the result (\ref{ad-field}). 

Let us now compute the renormalized self-energy, given by the finite part as $\epsilon \to 0$. The finite part can also be 
computed in terms of the structure function by using the general formula (\ref{finite-part}). We first note that, from \eqref{anomalous_F0} and $\beta_\lambda^{(0)}(\lambda) = -\lambda^2$, we have
\be
-\int_0^\lambda {F_0(-u) \over u} \rd u=- \int_0^\lambda {\gamma^{(1)}(u) \over \beta^{(0)}_\lambda (u)} \rd u. 
\ee
Therefore, in this case, \eqref{finite-part} reads
\be
\label{sigmap-r}
 \Sigma_{p,R}^{{\rm P},1}=- \int_0^\lambda {\gamma^{(1)}(u) \over \beta^{(0)}_\lambda (u)} \rd u- F_{0,1} \lambda+ \sum_{m \ge 2} (m-1)! F_m (0) \lambda^m. 
\ee
From the structure function (\ref{fxy}), we find 
\be
\label{f-combi}
-F_{0,1} \lambda=-{\lambda \over 4}, \qquad \sum_{m \ge 2}  F_m (0) y^m= -{y \over 2(2-y)}+{y \over 4}, 
\ee
where we have set $p^2=-\mu^2$ in (\ref{fxy}) and therefore the coupling constant appearing above is $\lambda= \lambda(p)$. 
Since $F_{0,1}= -F_{1,0}$, we obtain 
\be
\label{sigmaP}
 \Sigma_{p,R}^{{\rm P},1}=- \int_0^{\lambda(p)} {\gamma^{(1)}(u) \over \beta^{(0)}_\lambda (u)} \rd u-\sum_{m \ge 1} {(m-1)! \over 2^{m+1}} (\lambda(p))^m. 
\ee
To compare this result with (\ref{ts-sigma}) we have to take into account the change of scheme, from SM to $\overline{\text{MS}}$. 
The coupling constants $\lambda$ appearing in both expressions are the same, up to this order in 
$1/N$. In addition we have to take into account the factor $\zeta(\lambda)$ determined in (\ref{zeta-1N}). 
By using this result, we find that the relation between the self-energy $\Sigma^{\rm SM}_p$ in the SM scheme 
and the self-energy $\Sigma_{p,R}$ in the $\overline{\text{MS}}$ scheme, at leading order in the $1/N$ expansion, is given by 
\be
\label{correction}
\Sigma^{\rm SM}_p=  \Sigma_{p,R}+ \int_0^\lambda {\gamma^{(1)}(u) \over \beta^{(0)}_\lambda (u)} \rd u+ \CO\bigl(N^{-1}\bigr). 
\ee
Since the power series in the second term of the r.h.s. of (\ref{sigmaP}) is 
nothing but $-\Phi_0(\lambda)$, we conclude that the 
perturbative calculation of the self-energy in the $\overline{\text{MS}}$ scheme gives precisely the perturbative part of the 
trans-series (\ref{ts-sigma}), after taking into account the correction (\ref{correction}).  

Let us mention that the factorially divergent perturbative series (\ref{sigmaP}) is perhaps the simplest example of an IR renormalon in an 
asymptotically free theory. As is well-known, IR renormalons are smoking guns for non-perturbative corrections 
due to condensates. We will see later on in this paper that in the case of the renormalon (\ref{sigmaP}), the corresponding condensate is the 
four-quark condensate\footnote{In section IV of the original paper by Gross and Neveu \cite{gross-neveu} they consider the self-energy for the theory in which fermions have a mass term of the form (\ref{fermion-mass}). They show that the perturbative expansion of this quantity is factorially divergent, and this is the first appearance of a renormalon in the QFT literature. However their example is not an IR renormalon, but an UV renormalon.}.  

\subsection{General aspects of perturbation theory with condensates}

We would like now to calculate the power corrections appearing in (\ref{ts-sigma}) within the SVZ approach. The basic idea in this approach is to use 
the OPE for the operators appearing in the two-point function. In our case the OPE reads, schematically, 
\be
\label{ope}
\overline \bpsi (x) \bpsi(0)= C(x) {\bf 1}+ 
C_{\overline \psi \psi} (x) [\overline \bpsi (0) \bpsi (0)]+ C_{K} (x)[K(0)] + C_{V} (x)[ V(0)] + \cdots,
\ee
where we have included the operators of dimension one and two. One further assumes 
that the different operators appearing in the OPE 
have non-vanishing vevs, also called vacuum condensates. In the GN model this is expected to be so, as it can 
seen for example in the large $N$ solution of the model. Indeed, it is elementary to show that, in the $\overline{\text{MS}}$ scheme, the 
solution of the gap equation (\ref{gap}) for $\sigma_c=m_0$ is 
\be
\sigma_c^2= \Lambda^2, 
\ee
where we used the normalization for this field in (\ref{seff}). On the other hand, $\sigma$ can be integrated out and 
is given by
\be
\sigma= g_0 \overline \bpsi \bpsi, 
\ee
therefore we have
\be
\label{fermi-condensate}
 \langle [\overline \bpsi \bpsi] \rangle_c \approx -{N \over \pi \lambda} \Lambda
 \ee
at large $N$. The operator $[V]$ appearing in the Lagrangian should also have a non-trivial vev, by large $N$ factorization, and we expect
\be
\label{4q}
\langle [V]\rangle_c \approx {\pi \lambda \over N}  \langle [\overline \bpsi \bpsi ]\rangle_c^2 
\ee
at large $N$. In addition, due to (\ref{EOM}), we also expect 
\be
\label{KV}
\langle [K]\rangle_c = - \langle [V]\rangle_c. 
\ee
Therefore, we will assume that all the operators appearing in the OPE (\ref{ope}) lead to non-trivial vacuum condensates, and we will assume 
that these condensates satisfy the properties (\ref{4q}), (\ref{KV}), as they follow from large $N$ factorization and basic principles. 

Although we have written the OPE (\ref{ope}) in position space for pedagogical purposes, we will always 
work in momentum space, where the OPE is valid at large momentum. This is the standard setting for QCD 
sum rule calculations. The relation between the OPE in momentum and in position space is not completely 
straightforward, since some terms which appear in position space do not appear in momentum 
space (see e.g. the discussion at the beginning of \cite{nsvz-tr}), but we will not deal with these issues in this paper. 

In order to make contact with the trans-series (\ref{ts-sigma}) we need the precise relation 
between vacuum condensates and the dynamically generated scale $\Lambda$, which can 
be in turn related to the mass gap through (\ref{m-lambda}). This relation follows from general 
principles, since the vevs of composite operators have to satisfy 
the Callan--Symanzik equation
\be
\label{matrix-cs}
\left[ \delta_{ij}  \left(\mu {\partial \over \partial \mu}+ \beta(g) {\partial \over \partial g}  \right)+ \gamma_{ij} \right] \langle [\CO_j]\rangle_c=0, 
\ee
in the general case of operator mixing. If there is a single operator $\CO$ of dimension $d$, with anomalous dimension $\gamma_\CO$,  the solution of the equation (\ref{matrix-cs}) is
\be
\label{sol-nomix}
\langle [\CO] \rangle_c=\xi \Lambda^d \exp \left(- \int^{g(\mu)} {\gamma_\CO(u) \over\beta (u)}  \rd u \right), 
\ee
where $\xi $ is an overall constant which might depend on the parameters of the theory, like $N$. 
Eq.~(\ref{sol-nomix}) applies in particular to the operator $\overline \bpsi \bpsi$, which does not mix. In this case we have $\gamma_{\overline \psi \psi}=- \gamma_m$. By using the explicit expressions (\ref{gamma0}), (\ref{gammafirst}) for the mass anomalous dimension, one finds
\be
\label{cond-lambda}
\langle [ \overline \bpsi \bpsi ] \rangle_c = -N c(N) {\Lambda \over \pi \lambda} \exp\left(-{1\over N} \int^\lambda {\chi(u) \over u^2}  \rd u + \CO\bigl(N^{-2}\bigr) \right), 
\ee
where $\chi(u)$ is given in (\ref{chiF}). Its expansion around $u=0$ has the form 
\be
\chi(u)= \chi_1 u+ \CO(u^2), \qquad \chi_1={1\over 2}, 
\ee
 and the integral appearing in (\ref{cond-lambda}) has to be understood as  
\be
\label{con-clar}
 \int^\lambda {\chi(u) \over u^2}  \rd u = \chi_1 \log(\lambda) +  \int^\lambda_0 {\chi(u)-\chi_1 u \over u^2}  \rd u. 
 \ee
If we compare (\ref{cond-lambda}) with the result at large $N$ in (\ref{fermi-condensate}) we find that $c(N)$ has the $1/N$ expansion
\be
\label{cNas}
c(N)= 1 + {c_1 \over N}+\CO\bigl(N^{-2}\bigr).  
\ee
The choice of normalization in (\ref{cond-lambda}), with additional factors of $\pi$, $N$ was made so that $c(N) \approx 1$ at large $N$, as shown in (\ref{cNas}). The subleading terms in the expansion of $c(N)$ could be obtained by calculating $1/N$ corrections to the effective potential. 
%

In the case of the operators $K$, $V$, there is operator mixing and the matrix of anomalous dimensions is given in (\ref{gam-matrix}). We can reduce the system of equations to a single equation by taking into account (\ref{KV}), and solve it in the $1/N$ expansion. A simple calculation 
gives 
\be
\label{4quark-L}
\langle  [V] \rangle_c = N d(N) {\Lambda^2 \over \pi \lambda} \exp \left( {1\over N} \left({\beta^{(1)} (\lambda) \over \lambda^2}-1 \right) +\CO\bigl(N^{-2}\bigr)\right), 
\ee
where 
\be
\label{dNexp}
d(N)= 1+ {d_1 \over N} + \CO\bigl(N^{-2}\bigr). 
\ee
The fact that $d(N) \approx 1$ at large $N$ is a consequence of the large $N$ factorization (\ref{4q}). 

With the above results, we can calculate the value of the condensates 
appearing in the OPE (\ref{ope}), up to the non-perturbative 
functions $c(N)$, $d(N)$. The Wilson coefficients can be determined in 
various ways. The most practical method is 
the following (see \cite{pascual-tarrach} for an excellent exposition in the context of QCD): one expands 
in series the interaction terms in the Lagrangian, as in standard perturbation theory. We then Wick-contract the elementary fields 
with the usual rules, except for the fields that will enter into the condensate. For example, to calculate $C_{\overline \psi \psi}$, corresponding 
to the two-quark condensate, two of the elementary fields shouldn't participate in the contraction, but form the vacuum condensate. 
The combinatorial possibilities for doing this expansion can be represented by Feynman diagrams in which the fields that form the condensate 
are represented by blobs. We will see plenty of examples of this procedure in the next section.  

In forming the condensates, we will encounter vevs of fermion operators which are not Lorentz scalars or $U(N)$ singlets. These vevs can be determined easily by imposing Lorentz and $U(N)$ invariance. For example, for a general bilinear in fermions, we have
\be
\label{two-cond}
\langle \psi^\mu_i (0) \overline \psi^\nu_j (0) \rangle_c= -{\delta_{ij} \delta^{\mu \nu}\over 2 N} \langle \overline \bpsi \bpsi \rangle_c, 
\ee
while for the operators related to the Lagrangian operators $V$, $K$ we have
\be
\label{K0cond}
\ba
g_0 \langle \psi_i^\alpha (0) \overline \psi^\beta_j (0)  \psi_k^\mu (0) \overline  \psi^\nu_l  (0)\rangle_c&= \left(\delta^{\alpha \beta} \delta^{\mu \nu}\delta_{ij} \delta_{kl}-
\delta^{\alpha \nu} \delta^{\mu \beta}\delta_{il} \delta_{jk}\right) {\langle V \rangle_c \over  2N (2N-1)}, \\
\langle \partial_\mu \psi_i^\alpha (0) \overline \psi_j ^\beta (0) \rangle_c&= {\ri \over 2 d N} (\gamma_\mu)^{\alpha \beta} \delta_{ij} \langle K \rangle_c. 
\ea 
\ee
We note that we use a dimensional regularization in which the dimension of space-time is $d$, but the dimension of the 
Dirac spinor space is $2^{\lfloor d/2 \rfloor} = 2$. To relate the vevs of the bare operators to the finite vevs of the 
renormalized operators we use the renormalization constants in (\ref{bil-ren}), 
(\ref{Zmatrix}). For the operators $K$, $V$, using (\ref{EOM}), we find
\be
\langle V \rangle_c=-\langle K \rangle_c= \left( 1- {\beta_\lambda(\lambda) \over \epsilon \lambda} \right)  \langle [V] \rangle_c.
\label{V-renormalization}
\ee

\subsection{Trans-series for the two-quark condensate}

Let us now calculate the contribution to the self-energy of the trans-series associated to the two-quark condensate $\langle \overline \bpsi \bpsi \rangle_c$. 
Similar calculations in QCD can be found in \cite{politzer,pascual-der, elias}. To illustrate the OPE method with 
condensates, we will first consider in some detail the contribution at leading order in the coupling constant. A pedagogical exposition of the method in the case of QCD can be found in \cite{pascual-tarrach}. 

\begin{figure}[!ht]
\centering
\subcaptionbox{\label{condensatefirsts-fig-1}}{\includegraphics[height=3cm,trim={0 1.35cm 23cm 1.75cm},clip]{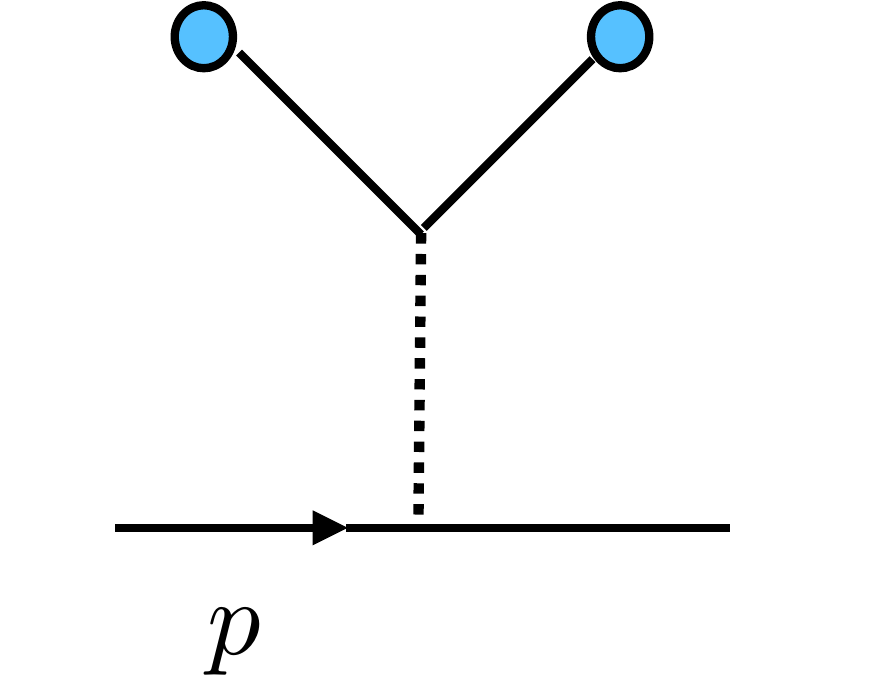}}
\subcaptionbox{\label{condensatefirsts-fig-2}}{\includegraphics[height=3cm,trim={19.5cm 1.6cm 0 1.5cm},clip]{first-order-condensates.pdf}}
\caption{The two diagrams that contribute to the two-quark condensate correction to the two-point function, 
at leading order in $g_0$.}
\label{condensatefirsts-fig}
\end{figure}

It is convenient to start the calculation in position space, and only later Fourier transform into momentum space. We recall the Wick contractions are defined as
\be
\label{tau-con}
\wick{\c1{\sigma} (x) \c1{\sigma} (y)= -\ri \delta(x-y)}
\ee
and
\be
\wick{\c1{\psi}^\mu_i (x)\c1{\overline \psi}_j^\nu (y)= \delta_{ij} S_0^{\mu \nu}(x-y)}, 
\ee
where 
\be
S_0^{\mu \nu}(x)= \int  {\rd^d k \over (2 \pi)^d} \re^{-\ri k x} \left[{\ri \over \slashed k}\right]^{\mu \nu}.
\ee
To compute the two-point function at order $g_0$, we bring down two factors 
of the interaction term in the Lagrangian. This yields the following expression
\be
- g_0 \langle  \psi^\mu_i (x) \overline \psi^\nu_j (0) \int \rd^d y_1 \rd^d y_2 \, \overline \psi^\alpha_m (y_1) \psi^\alpha_m (y_1) \sigma(y_1) 
\overline \psi^\beta_n (y_2) \psi^\beta_n (y_2) \sigma(y_2) \rangle, 
\ee
where we have to perform Wick contractions, but leaving a two-quark pair uncontracted to form the condensate. There are two ways of doing this. In the first way, we contract 
both external legs to the same vertex. One set of contractions is 
\be
\langle \wick{\c1{\psi}^\mu_i (x) \c2{\overline \psi}_j^\nu (0) \c1{\overline \psi}_m^\alpha (y_1) 
\c2{\psi}_m^\alpha (y_1)  \c1{\sigma} (y_1) \overline \psi^\nu_b (y_2) \psi^\nu_b (y_2) \c1{\sigma} (y_2)
 } \rangle
 \ee
 and there is another, equivalent one obtained by exchange of the vertices $y_1$ and $y_2$. This  type of contractions can be 
 represented by the diagram in \figref{condensatefirsts-fig-1}. Note that the two-quark pair which leads to the vacuum condensate is traced over. After performing a Fourier transform into momentum space and doing the integrals in the spacetime variables, we obtain 
\be
-{\ri g_0 \over p^2} \langle \overline \bpsi \bpsi\rangle_c.
\label{diagram2}
\ee
This is a contribution of order one in the $1/N$ expansion and, after renormalization, it will give the term $m$ in the trans-series of the self-energy \eqref{ts-sigma}. 

In the second type of contributions, we contract each external leg with a different vertex. This leads to contractions of the form
\be
\langle  \wick{\c1{\psi}^\mu_i (x) \c2{\overline \psi}_j^\nu (0) \c1{\overline \psi}_m^\alpha (y_1) \psi_m^\alpha (y_1)  \c3{\sigma} (y_1) \overline \psi^\beta_b (y_2) \c2{\psi}^\beta_b (y_2) \c3{\sigma}(y_2)
 } \rangle
 \ee
as well as a similar one obtained by exchanging the vertices. They can be represented by the diagram in \figref{condensatefirsts-fig-2}.
The vev of the product $\psi_m^\alpha (y_1)  \overline \psi^\beta_b (y_2)$ leads to a condensate and, due to the delta function $\delta(y_1-y_2)$ coming from (\ref{tau-con}), both fields are at the same point. After using (\ref{two-cond}), we find that the contribution of the diagram in \figref{condensatefirsts-fig-2} to the two-point function is simply
\be
\label{diagram1}
{ \ri g_0  \over 2N} {1\over p^2} \langle \overline \bpsi \bpsi\rangle_c. 
\ee
A general principle to retain from this calculation is that condensates in which both 
quarks come from the same interaction vertex have a relative factor 
of $N$, as compared to condensates where the fermions are from different vertices. This is important when 
taking into account large $N$ counting. 

After multiplying the diagrams in \figref{condensatefirsts-fig} by $\ri$ and removing their external legs, we find that the total contribution of the two-quark trans-series to 
the self-energy, at first order in the coupling, is 
\be
\label{leading-self}
-g_0 \left(1-{1\over 2N} \right) \langle \overline \bpsi \bpsi \rangle_c. 
\ee
Let us mention that the diagram in \figref{condensatefirsts-fig-2} has a counterpart in the calculation of the 
two-quark condensate correction to the quark propagator in QCD \cite{politzer, pascual-der}, in which the $\sigma$ propagator is replaced with a gluon.

\begin{figure}[!ht]
\centering
\subcaptionbox{\label{nocontribute-fig-1}}{\includegraphics[height=2.3cm,trim={0 3.425cm 23.5cm 3.225cm},clip]{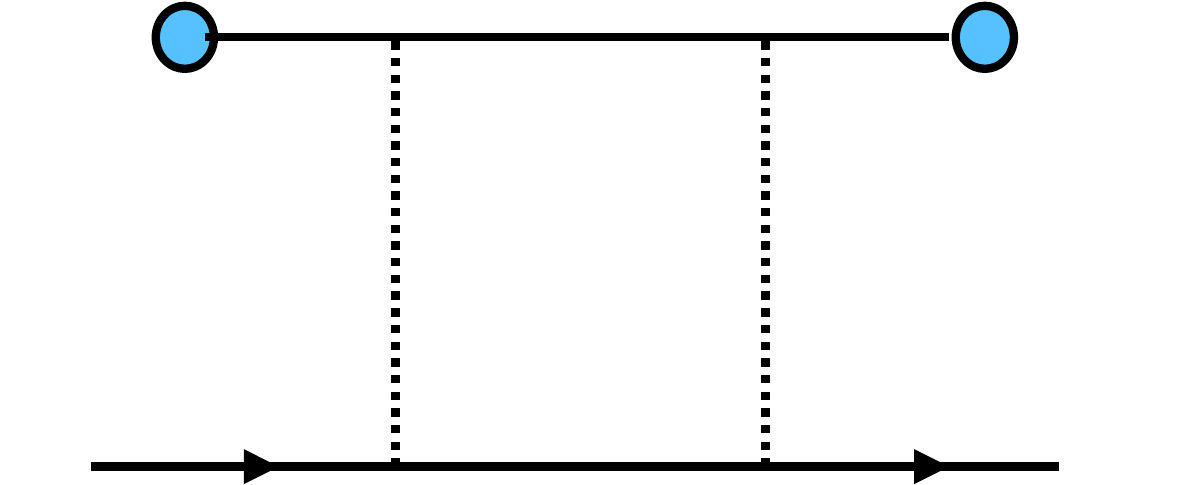}}
\subcaptionbox{\label{nocontribute-fig-2}}{\includegraphics[height=2.3cm,trim={23.5cm 3.6cm 0 2.8cm},clip]{nocontribute-2quark.pdf}}
\caption{These diagrams are of order $1/N$, but they do not contribute to the two-quark condensate trans-series. Diagram \subref{nocontribute-fig-1} vanishes since it is proportional to $\tr \, \gamma^\mu=0$, and \subref{nocontribute-fig-2} involves propagators at zero momentum.}
\label{nocontribute-fig}
\end{figure}

After this pedagogical exercise, let us consider the diagrams which give the trans-series 
$\Phi_{1}(\lambda)$ in the self-energy (\ref{ts-sigma}). 
In order to proceed, we recall that in calculating the self-energy in 
conventional perturbation theory, one considers only 
one-particle irreducible diagrams, i.e. diagrams that can not be divided into 
subdiagrams by cutting one fermionic internal line. 
The underlying reason is that reducible diagrams factorize, therefore their contribution 
can be obtained from 
their subdiagrams. In the presence of condensates one has to be careful, 
since diagrams that look reducible in the 
conventional sense do not factorize. Let us then reconsider 
the relationship between the two-point function and the 
self-energy when one includes non-perturbative corrections coming from condensates.

As it is clear from (\ref{ts-sigma}), the self-energy up to order $1/N$ is a trans-series. The inverse, renormalized 
two-point function can then be written as 
\be
\ri S^{-1}_R (p)=  \slashed p -{1\over N} \slashed p \Sigma_{p,R}^{\rm P} - 
\CC\Sigma_{m,R}^{\rm NP} -{\CC^2 \over N}  \slashed p \Sigma_{p, R}^{\rm NP} +\CO\bigl(\CC^3\bigr).  
\ee
In this equation, we have introduced a trans-series parameter $\CC$ to keep track of the powers of $m$, the superscripts ${\rm P}$, ${\rm NP}$ refer to 
perturbative and non-perturbative contributions, respectively, and the subscript $R$ stands for renormalized. The self-energies appearing in this equation have $1/N$ expansions with the structure 
\be
\label{sigmaRdec}
\ba
\Sigma_{p,R}^{\rm P}&=\sum_{j \ge 1} \Sigma_{p,R}^{{\rm P},j} N^{-j+1}, \\
\Sigma_{m,R}^{\rm NP}&=\sum_{j \ge 0} \Sigma_{m,R}^{{\rm NP},j} N^{-j},\\
\Sigma_{p,R}^{\rm NP}&=\sum_{j \ge 1} \Sigma_{p,R}^{{\rm NP},j} N^{-j+1}. 
\ea
\ee
An important remark is that, in doing the $1/N$ expansions of the self-energies computed diagramatically, 
we keep the condensates themselves fixed, i.e. we do not expand them as in 
(\ref{cond-lambda}) or (\ref{4quark-L}). This is the natural $1/N$ counting when working with diagrams with condensates. 
We now expand $S_R (p)$ in $\CC$ and $1/N$ to obtain 
\be 
\label{self-exp-1}
S_R (p)= {\ri \over \slashed p}  \left\{ 1+ { 1 \over N}\Sigma_{p,R}^{{\rm P},1} + {\CC \over \slashed p} \left( \Sigma_{m,R}^{{\rm NP}, 0}+ {1\over N} \Sigma_{m,R}^{{\rm NP},  1}\right) + {2 \CC \over  N \slashed p}  \Sigma_{m,R}^{{\rm NP}, 0}\Sigma_{p,R}^{{\rm P},1}
+\CO\bigl(\CC^2, N^{-2}\bigr) \right\}. 
\ee
The second term inside the brackets is the perturbative piece calculated in \eqref{sigmaP}, 
and $\Sigma_{m,R}^{{\rm NP}, 0}$ can be obtained from \eqref{diagram2}. The 
fourth term appearing inside the brackets in the r.h.s. factorizes into a perturbative piece and a non-perturbative piece, and as we will see it corresponds to a reducible diagram.

Let us now consider the diagrams that contribute to (\ref{self-exp-1}). They have to be of order $1/N$, but incorporate all 
loops. In conventional perturbation theory, such diagrams are obtained by inserting chain bubbles, and the same principle holds in the 
case of perturbation theory with condensates. There are however various diagrams that have the right $1/N$ counting but vanish in dimensional 
regularization, or vanish because they involve a trace of an odd number of gamma matrices. An example is the 
diagram in \figref{nocontribute-fig-1}. There is another type of diagrams that do not contribute: condensates are essentially 
zero-momentum insertions and they can lead to diagrams in which we have propagators at zero momentum. These diagrams have 
to be discarded \cite{pascual-tarrach}. The diagram in \figref{nocontribute-fig-2} is an example of this. We note however that 
the diagram in \figref{nocontribute-fig-1} turns out to contribute to the four-quark condensate correction, as we will explain in section \ref{sec_4quark}.

\begin{figure}[!ht]
\centering
\includegraphics[height=2.5cm,trim={0 3cm 0 4cm},clip]{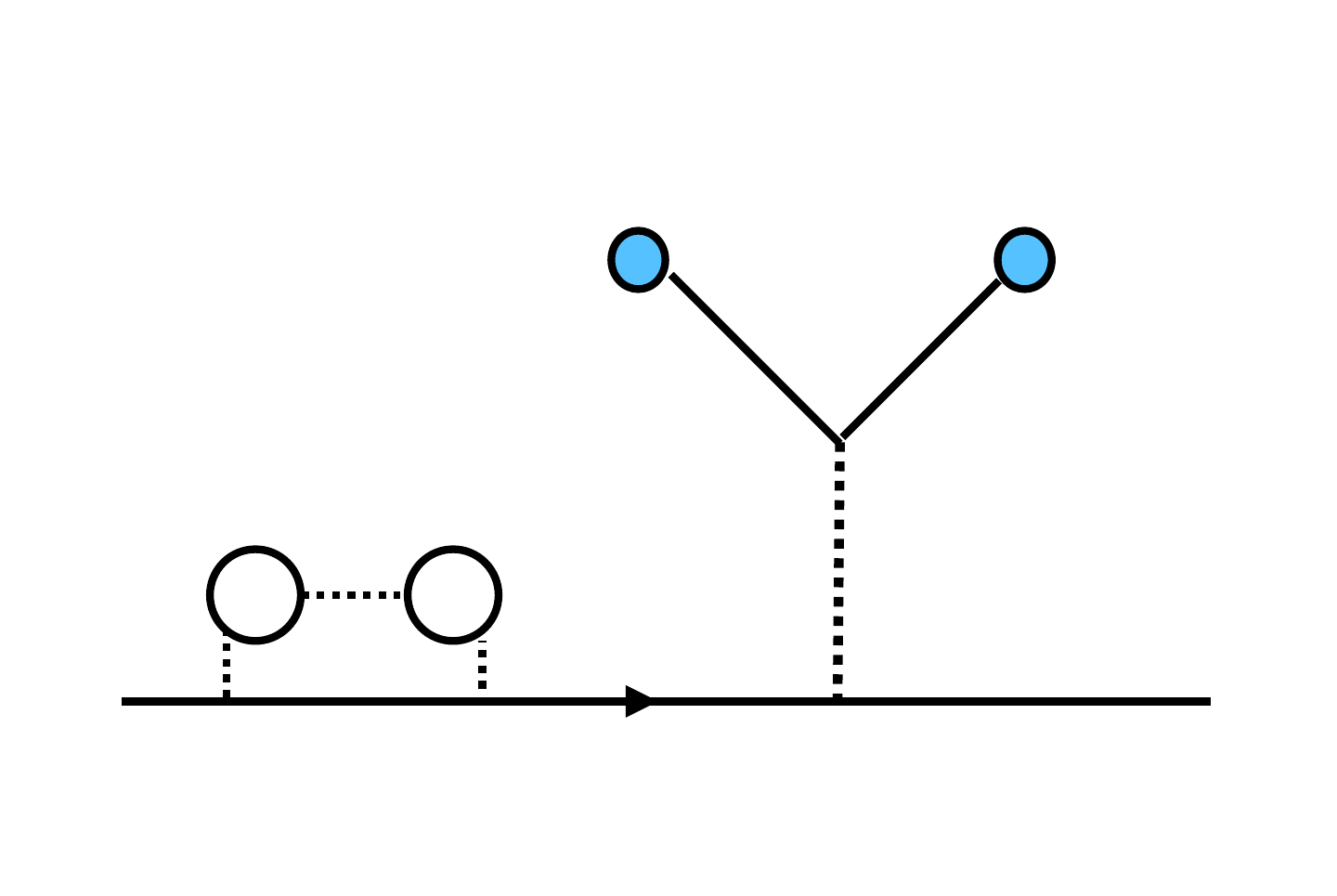}
\caption{A reducible diagram.}
\label{condensateN2-fig}
\end{figure}

Among the non-vanishing diagrams, one finds \figref{condensateN2-fig}. Its contribution factorizes into the contribution of the 
perturbative diagram of \figref{chain-bubble-fig}, which is of order $1/N$, and the contribution of the diagram in 
\figref{condensatefirsts-fig-1}. This is the reducible diagram that corresponds to the fourth term inside the bracket in 
(\ref{self-exp-1}), as we anticipated above. It does not contribute to the 
self-energy. 

\begin{figure}[!ht]
\centering
\subcaptionbox{\label{condensateN-fig-1}}{\includegraphics[height=3cm,trim={0 1.55cm 22cm 1.55cm},clip]{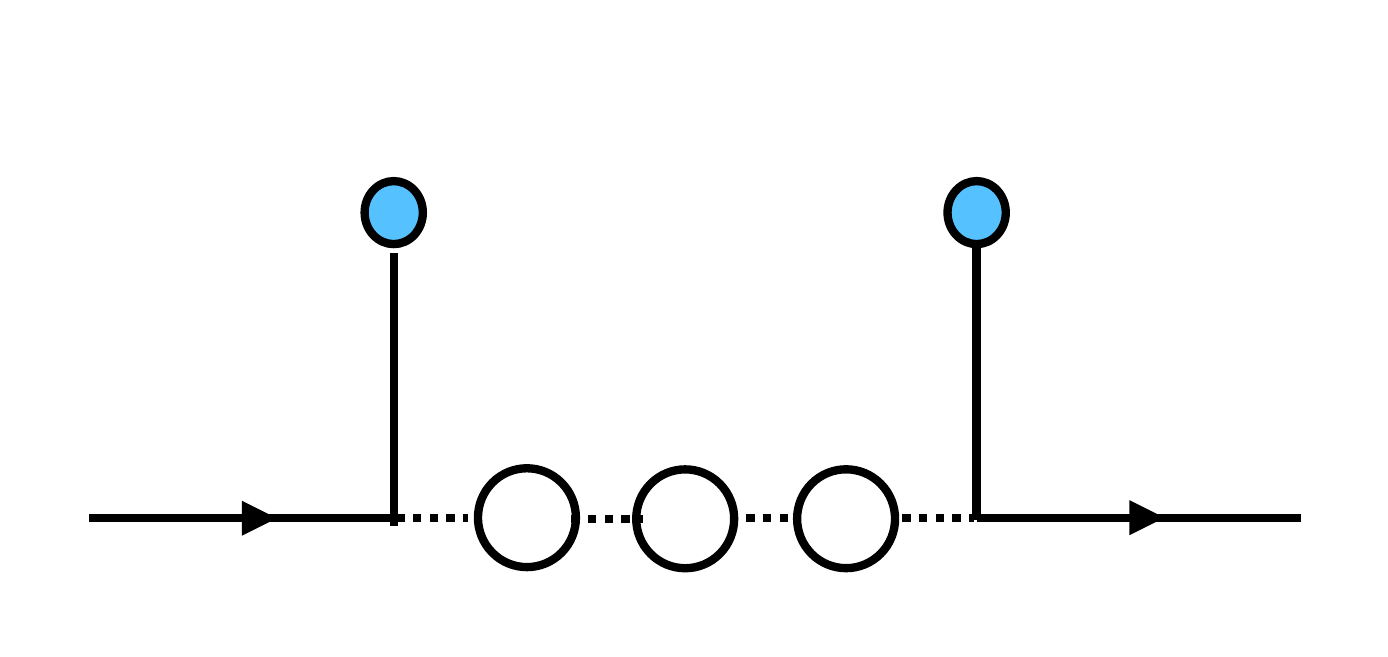}}
\subcaptionbox{\label{condensateN-fig-2}}{\includegraphics[height=3cm,trim={26.5cm 1.4cm 0 1.7cm},clip]{twoquark-N-fig.pdf}}
\caption{Irreducible diagrams that contribute to the two-quark condensate trans-series at order $1/N$, and all loops.}
\label{condensateN-fig}
\end{figure}

Let us then consider diagrams which do not factorize. We will call such diagrams irreducible. They 
are shown in \figref{condensateN-fig} (in the drawings we show only insertions of two or three bubbles, 
but of course one should consider insertions of an arbitrary number of bubbles). The diagram in \figref{condensateN-fig-1} is obtained by 
inserting the bubble chain \eqref{chain-bubble} inside in \figref{condensatefirsts-fig-2}, and no additional integration 
is needed. Its contribution to the self-energy is
\be
\label{chain-tree}
{  \pi \lambda_0 \over 2N^2 } \langle \overline \bpsi \bpsi\rangle_c \sum_{n \ge 1} \left(\ri \Pi (p^2)\right)^n  \left( \pi \lambda_0 \right)^n,  
\ee
where $n$ is the number of polarization loops inserted.\footnote{In this equation, we start the sum at $n=1$, since the term $n=0$ will be already accounted for when we add \eqref{leading-self}.} Notice that, when $n \ge 1$, the fermion and antifermion fields in the condensate are no longer at the same point. In this case, we have to expand
\be
\label{2-expand}
\ba
\langle \psi_m^\alpha (y_1)\overline \psi^\beta_n (y_2) \rangle_c&= \langle \psi_m^\alpha (0)\overline \psi^\beta_n (0) \rangle_c+ (y_1- y_2)^{\mu}  \langle \partial_\mu \psi_m^\alpha (0)\overline \psi^\beta_n (0) \rangle_c+ \cdots \\
&= -{1\over 2 N} \delta_{mn} \delta^{\alpha \beta} \langle\overline \bpsi \bpsi\rangle_c+ \cdots.
\ea
\ee
In calculating the contribution to the two-quark condensate we only retain the first term in the r.h.s. of the first line, but the derivative term will contribute to the four-quark condensate, as we will see in the next section.

Let us now consider the diagram in \figref{condensateN-fig-2}. A straightforward calculation gives the following contribution to the 
self-energy:
\be
\label{bubble-int}
{\pi \lambda_0  \over N^2 } \langle \overline \bpsi \bpsi \rangle_c \sum_{n \ge 1}  \ri \CI_1(n) \lambda_0^{n}, 
\ee
where $\CI_1(n)$ is the loop integral 
\be
\CI_1(n)=\pi^n \int {\rd^d q \over (2 \pi)^d} {  \left( \ri \Pi (q^2) \right)^{n-1} \over (p+q)^2}. 
\ee
This integral can be evaluated in dimensional regularization using (\ref{gen-loop}):
\be
\CI_1(n)=-{\ri \over 4} \left( -{p^2 \over 4 \pi} \right)^{-\epsilon n/2} {\Gamma\left(-{\epsilon \over 2} \right) \Gamma \left( {n \epsilon \over 2} \right) \Gamma\left(1-{n \epsilon \over 2} \right)  \over \Gamma \left( {(n-1) \epsilon \over 2} \right)\Gamma \left( 1-{(n+1) \epsilon \over 2} \right)} 
\left[ -{1\over 2} \frac{\Gamma\left(1-\frac{\epsilon}{2}\right)\Gamma\left(1+\frac{\epsilon}{2}\right)\Gamma\left(-\frac{\epsilon}{2}\right)}{\Gamma(1-\epsilon)} \right]^{n-1}. 
\ee

We can now add the contributions of the two diagrams, together with (\ref{leading-self}), to obtain the bare, non-perturbative correction to the 
term $\Sigma_m$ in the self-energy: 
\be
\Sigma^{\rm NP}_m=- { \pi \lambda_0 \over N }\langle \overline \bpsi \bpsi \rangle_c \Biggl\{  1-{1\over 2 N} - {1 \over N} \sum_{n \ge 1} \left(  \ri \CI_1(n) +{1\over 2}(\ri \pi \Pi)^n \right) \lambda_0^n + \CO\bigl(N^{-2}\bigr) \Biggr\}.
\ee
There are three sources of renormalization in this quantity: renormalization of the coupling constant, 
renormalization of the self-energy, and renormalization of the composite operator appearing in the condensate 
(see \cite{huangs} for useful remarks on renormalization 
of fermion propagators in perturbation theory with condensates). 
The renormalization of the self-energy is done as 
in the perturbative case, and it just follows by multiplying the 
inverse two-point function by $Z_\psi$.  The renormalization of the 
composite operator follows from (\ref{Zpp}):
\be
\langle \overline \bpsi \bpsi \rangle_c= Z_m^{-1} \langle [\overline \bpsi \bpsi ]\rangle_c. 
\ee
The renormalized result is
\be 
\label{self-NP}
\Sigma^{\rm NP}_{m, R} =-{ \pi  \lambda  \over N }\langle [\overline \bpsi \bpsi ]\rangle_c  {Z_\lambda  \over Z_m} \Biggl\{  1-{1\over 2 N}+ {1\over N} Z_\psi^{(1)} - {1 \over N} \sum_{n \ge 1} \left(  \ri \CI_1(n) +{1\over 2}(\ri \pi \Pi)^n \right) \lambda_0^n + \CO\bigl(N^{-2}\bigr) \Biggr\}. 
\ee
Let us note that the sign of the renormalization constant $Z_\psi^{(1)}$ is the opposite one to 
what is found for the perturbative part in (\ref{self-p-ren}). This can be seen by comparing (\ref{self-exp-1}) to the diagrammatic expansion, or 
simply by noting that $\slashed p$ and $m$ have opposite signs in the inverse propagator. The renormalization constants $Z_\lambda$, $Z_m$ have a $1/N$ expansion 
\be
\label{1NZs}
\ba
Z_\lambda= Z^{(0)}_\lambda \left(1+ {1\over N} \widehat Z^{(1)}_\lambda+ \CO\bigl(N^{-2}\bigr) \right), \\
Z_m= Z^{(0)}_m \left(1+ {1\over N} \widehat Z^{(1)}_m+ \CO\bigl(N^{-2}\bigr) \right), 
\ea
\ee
and their first terms are equal:
\be
Z^{(0)}_\lambda=Z^{(0)}_m. 
\ee
%
%
%
%
%
A first consistency check of (\ref{self-NP}) is that it is finite, i.e. that the divergences 
in the diagrams of \figref{condensateN-fig} cancel against the renormalization constants. 
To verify that, and to calculate the finite part, we will calculate the sum over $n$ appearing here by using the formalism 
explained in Appendix \ref{pmp-trick}. The structure function which calculates the sum
\be
- \sum_{n \ge 1} \left(  \ri \CI_1(n) +{1\over 2}(\ri \pi \Pi)^n \right)\lambda_0^n  
\ee
is given by
\begin{multline}
\label{hxy}
H(x,y)=-{1\over 2} \left( -{p^2 \over 4 \pi \nu^2} \right)^{-y/2}
\left[  \frac{\Gamma\left(1+\frac{x}{2}\right) \Gamma^2\left(1-\frac{x}{2}\right)}{\Gamma(1-x)} \right]^{y/x-1}\\
\times
\left[  {\Gamma \left(1+ {y \over 2}\right) \Gamma\left(1-{y \over 2} \right) \Gamma\left(-{x \over 2} \right)  \over \Gamma \left( {y-x \over 2} \right)\Gamma \left( 1-{y+x\over 2} \right)} - y \frac{\Gamma\left(1+\frac{x}{2}\right) \Gamma\left(1-\frac{x}{2}\right) \Gamma \left(-{x\over 2} \right)}{2\Gamma(1-x)}  \right].
\end{multline}
In particular, we have 
\be
H_0(\epsilon)=H (\epsilon,0)= {\chi(-\epsilon) \over \epsilon}- F_0(\epsilon).  
\ee
In this expression, $F_0(\epsilon)$ is the function defined in (\ref{F0}) and appearing in the calculation of 
the perturbative self-energy, and $\chi(\lambda)$ is the function (\ref{chiF}) in the mass anomalous dimension. Moreover, the $1/N$ expansion of the renormalization constants gives
\be
\label{zz-2}
\widehat Z^{(1)}_\lambda-  \widehat Z^{(1)}_m=   \int_0^\lambda \rd u {\chi(u) \over u (u+\epsilon)}. 
 \ee
Using (\ref{F0div}) and (\ref{divzpsi}), we obtain 
\be
\label{H0chi}
\left[ H_0(\epsilon) \log\left(1+ {\lambda \over \epsilon} \right) \right]_{\rm div}=   
- \widehat Z^{(1)}_\lambda +  \widehat Z^{(1)}_m - Z_\psi^{(1)}.
\ee
Now it is clear that all the divergent parts cancel in (\ref{self-NP}). In fact, one can use this calculation to determine 
the function $\chi(u)$, which essentially gives the mass anomalous dimension at NLO in the $1/N$ expansion. 

Let us now evaluate the finite part of the sum. As shown in (\ref{finite-part}), it has two pieces. One of them  
can be read from the structure function evaluated at $x=0$:
\be
H(0,y)= {y \over 4} \left( 2 \gamma_E  + \psi \left( 1-{y \over 2} \right) + \psi \left( {y \over 2} \right) \right). 
\ee
After expanding the above expression in powers of $y$, we can extract the coefficients $H_m(0)$ and obtain
\be
 -{1\over 2} +  \sum_{m \ge 1} (m-1)! H_m(0) \lambda^m =-{1\over 2} -\sum_{k \ge 1} {(2k)! \over 2^{2k+1}} \zeta(2k+1)\lambda^{2k+1}, 
\ee
where we have included the term $-1/2$ originating from the diagram in \figref{condensatefirsts-fig-2}. The other finite piece combines 
with the non-trivial power series in $\lambda$ which appears in (\ref{cond-lambda}), when 
we express the two-quark condensate in terms of $\Lambda$, to produce
\be
-\int^\lambda {\chi(u) \over u^2} \rd u -\int_0^\lambda {H_0(-u) -H_{0,0} \over u} \rd u = -{1\over 2} \log(\lambda) + \int_0^\lambda {\gamma^{(1)}(u) \over \beta^{(0)}_\lambda (u)} \rd u. 
\ee
We finally obtain 
\be
\label{mass-corr}
\ba
\Sigma^{\rm NP}_{m,R}=c(N) \Lambda & \Biggl\{ 1 -{1\over N} \Biggl({1\over 2} + {1\over 2} \log(\lambda) + \sum_{k \ge 1} {(2k)! \over 2^{2k+1}}\zeta(2k+1) \lambda^{2k+1} \Biggr)\\
& \qquad  + {1\over N} \int_0^\lambda {\gamma^{(1)}(u) \over\beta_\lambda^{(0)}(\lambda)} \rd u+ \CO\bigl(N^{-2}\bigr) \Biggr\}. 
\ea
\ee
The term in the second line is due to the change of scheme, since we have  
\be
\label{correction2}
m+ {m \over N} \Sigma^{{\rm SM}}_m=  \Sigma_{m,R} \left(1- \frac{1}{N}\int_0^\lambda {\gamma^{(1)}(u) \over \beta^{(0)}_\lambda (u)} \rd u 
+ \CO(N^{-2}) \right).   
\ee
Note that the correction involving the anomalous dimension of the field now comes with 
a minus sign, as compared to (\ref{correction}). We remove this scheme dependent term, write $\Lambda$ in terms of $m$ using the relation (\ref{m-lambda}), and expand the factor $c(N)$ using (\ref{cNas}). We finally obtain, from \eqref{mass-corr},
%
%
%
\be
-\Sigma^{{\text{SM}}}_m \sim 1-{\gamma_E \over 2} + \log(2) -c_1 + {1\over 2} \log(\lambda) + \sum_{k \ge 1} {(2k)! \over 2^{2k+1}} \zeta(2k+1) \lambda^{2k+1} + \mathcal{O}\bigl( N^{-1} \bigr).
\ee
Due to the second equation in (\ref{leadSM}), the series in the r.h.s. should be equal to the series 
$\Phi_{1}(\lambda)$ in (\ref{ts2}). This is indeed the case for the $\lambda$-dependent terms. 
Equality of the constant term fixes the value of $c_1$: 
\be
c_1= {1\over 2} + {3\over 2}  \log(2). 
\ee
The calculation above also fixes that $c(N) \approx 1$ at large $N$. We have already incorporated this information by using 
the large $N$ calculation of the condensate, 
but we could have obtained it from the comparison of our result (\ref{mass-corr}) with the explicit result for the trans-series. 

Let us note that our calculation of the two-quark condensate correction to the fermion self-energy
is conceptually very similar to what has been done for QCD in \cite{politzer,pascual-der, elias}. 
One of the goals of such a calculation in QCD is to dynamically generate a mass for the quarks 
out of the condensate. We know from the large $N$ 
analysis of \cite{gross-neveu} that this is indeed the case in the GN model: 
the non-zero vev of $\sigma$ gives simultaneously a vev to the two-quark condensate and a mass 
to the fermions. However, these two quantities are conceptually different (e.g. the first one is 
not RG invariant, while the second one is), and this difference makes itself manifest at next-to-leading order in the $1/N$ expansion.  
Our analysis of the self-energy shows in detail how the contribution of the two-quark condensate to the 
OPE of the fermion propagator generates a mass pole at large $N$ and an additional power correction at order $1/N$, 
in the manner intended in QCD. 

\subsection{Trans-series for the four-quark condensate}
\label{sec_4quark}

We will now consider the contribution of the four-quark condensate to the trans-series. In the calculation of the two-point function 
there are two possible sources for such condensate. First, we can leave two pairs of fermions uncontracted when expanding the action. 
This leads to factors of $\langle V \rangle_c$. Second, we can consider diagrams with a single pair of uncontracted fermions, like the ones we studied in the previous 
section, but at different locations. When we expand them as in (\ref{2-expand}) we will get factors of $\langle K \rangle_c$. Although 
this operator is strictly speaking not a four-quark operator, its vev gives the same contribution, but with an opposite sign, as we saw in (\ref{KV}). We will then refer to these contributions as also due to a four-quark condensate, by a slight abuse of language.

\begin{figure}[!ht]
\centering
\includegraphics[height=2.5cm,trim={0 2.2cm 0 2.5cm},clip]{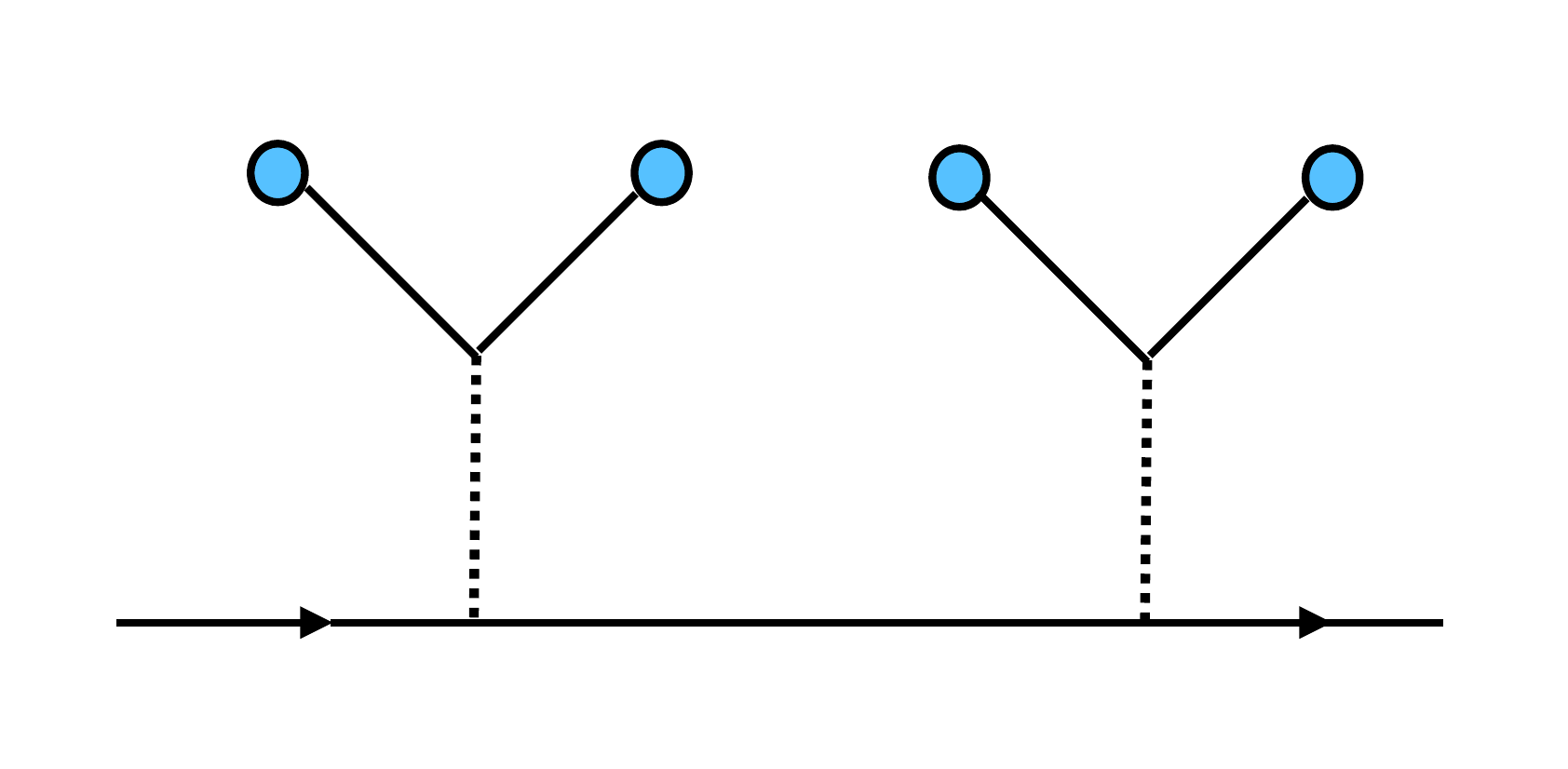}
\caption{Diagram of order $g_0^2$ and order $1$ in the $1/N$ expansion that contributes to the four-quark condensate correction to the two-point function.}
\label{4quarkL-fig}
\end{figure}

In order to proceed, we have to understand how to extract the self-energy from the two-point function, as we did in the 
case of the two-quark condensate. To see how this goes, 
let us consider the possible diagrams with a four-quark condensate which contribute to the two-point function. It is easy to see 
that there is a diagram, and only one, which gives a contribution of order one in the $1/N$ expansion, and shown in \figref{4quarkL-fig} (the condensates come from 
quark pairs in the same vertices, so they give a factor of $N^2$, and the diagram goes like $g_0^2 N^2$). A precise evaluation of the diagram gives 
\be
{g_0 \langle  V \rangle_c \over p^2}. 
\ee
Note that this does not factorize, so the diagram in \figref{4quarkL-fig} should not be regarded as a reducible diagram. The 
factorization only takes place at large $N$. In order to do the precise large $N$ counting, 
let us already renormalize the contribution of this diagram. For the composite operator $V$, we need the renormalization constant in \eqref{V-renormalization}:
\be
Z_{V} = 1-{\beta_\lambda(\lambda) \over \lambda \epsilon}= Z_V^{(0)} \left(1+ {1\over N} \widehat Z_{V}^{(1)} + \CO\bigl(N^{-2}\bigr) \right)
\ee
and we have 
\be
Z_V^{(0)} Z_\lambda^{(0)}=1, \qquad \widehat Z_{V}^{(1)}=- {\beta_\lambda^{(1)} (\lambda) \over  \lambda (\lambda+ \epsilon)}. 
\ee
Using in addition the renormalization constant of the coupling and its $1/N$ expansion in (\ref{1NZs}), we find 
\be
Z_\lambda Z_{V}{\pi \lambda \over N} \langle [V]\rangle_c= {\pi \lambda \over N} \langle [V]\rangle_c
 \left( 1+ {1\over N} \left( \widehat Z_\lambda^{(1)} + \widehat Z_{V}^{(1)} \right) + \cdots \right).
 \label{V-condensate-renormalization}
\ee
Note that we have the large $N$ scaling $\langle [ V]\rangle_c \sim N$, so 
the front factor is of order $1$. We will denote the renormalized 
two-quark and four-quark condensates by 
\be
\mathcal{T} = - \frac{\pi \lambda}{N} \langle [\overline{\bpsi} \bpsi]\rangle_c, \qquad \CF= {\pi \lambda \over N} \langle [V]\rangle_c. 
\ee
%
%
%
By the large $N$ factorization of \eqref{4q}, one has 
\be
\label{ft2}
\CF- \CT^2 = \CO\bigl(N^{-1}\bigr). 
\ee
We also note that 
\be
\Sigma_{m,R}^{{\rm NP},0}= \CT. 
\ee

\begin{figure}[!ht]
\centering
\subcaptionbox{\label{Nreducible-fig-3}}{\includegraphics[height=3cm,trim={3.75cm 6cm 34.25cm 20.5cm},clip]{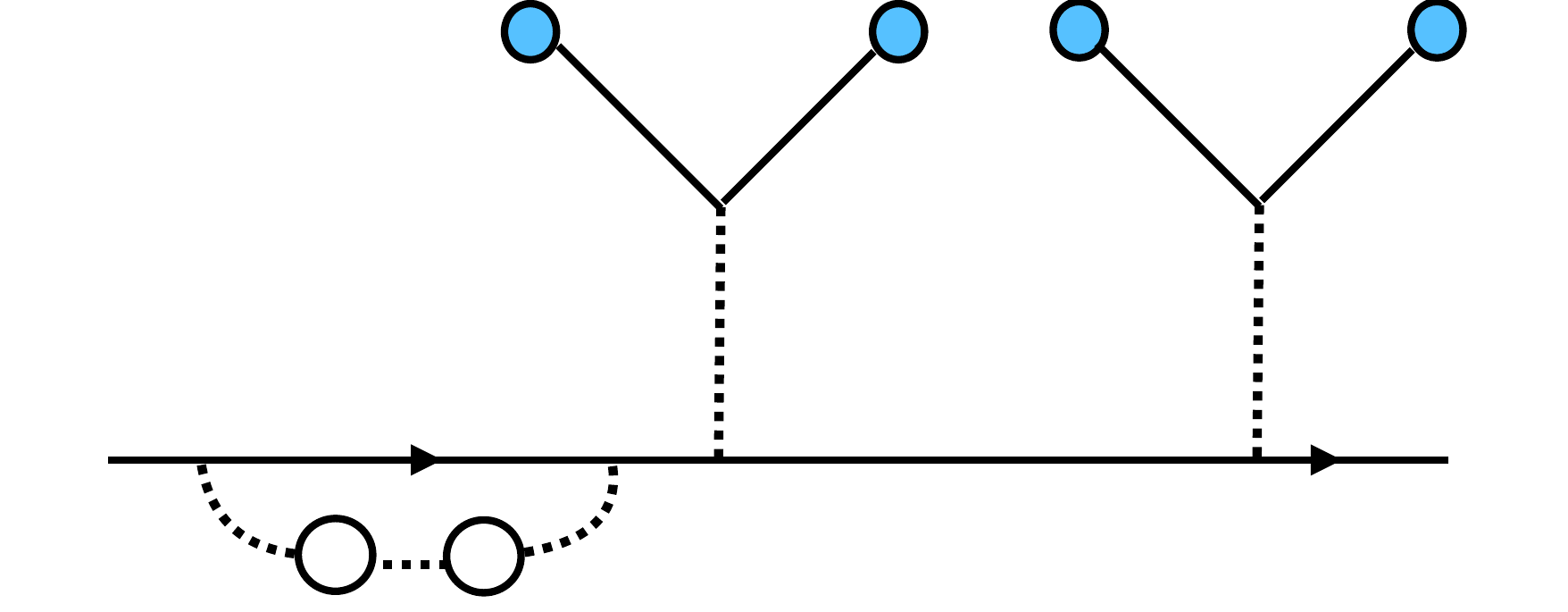}}
\subcaptionbox{\label{Nreducible-fig-4}}{\includegraphics[height=3cm,trim={34.25cm 6cm 4.75cm 20.5cm},clip]{largeN4quark-fig.pdf}}
\subcaptionbox{\label{Nreducible-fig-1}}{\includegraphics[height=3cm,trim={3cm 23.2cm 35cm 3.3cm},clip]{largeN4quark-fig.pdf}}
\subcaptionbox{\label{Nreducible-fig-2}}{\includegraphics[height=3cm,trim={35cm 23.5cm 4cm 3cm},clip]{largeN4quark-fig.pdf}}\\[8mm]
\caption{Diagrams with four-quark condensates that contribute to the two-point function but not to the self-energy, at order $1/N$. One should include as well the diagrams 
obtained by left-right reflection.}
\label{Nreducible-fig}
\end{figure}

In addition to the diagram of \figref{4quarkL-fig}, which is of order one at large $N$, 
we find many other diagrams of order $1/N$ in the calculation of the four-quark condensate corrections. Among these, there are 
the diagrams shown in \figref{Nreducible-fig} (together with their right-left reflections). These diagrams 
do not contribute to the self-energy, up to order $1/N$. 
To see this, we note that the term of order $\CC^2$ in the expansion of the renormalized two-point function (\ref{self-exp-1}) is, up to order $1/N$, 
 \be
{1\over N}  \Sigma_{p,R}^{{\rm NP},1}+{1\over p^2} \left( \CT^2  + {2 \over N} \CT \Sigma_{m,R}^{{\rm NP},  1} \right) + {3 \over p^2 N} \CT^2 \Sigma_{p,R}^{\rm P,1}.
\ee
To compare this expression with the diagrammtic computation, we still need to include the field renormalization constant and write renormalized quantities together with their original divergences. This yields
\begin{multline}
\frac{1}{N} \Sigma_{p,R}^{\text{NP},1} + \frac{1}{p^2} \CT^2 + \frac{2}{p^2 N} \CT^2 \left( \widehat{Z}_\lambda^{(1)} - \widehat{Z}_m^{(1)} \right)\\
+ \frac{3}{p^2 N} \CT^2 \left(\Sigma_{p,R}^{\text{P},1} + Z_\psi^{(1)}\right)
+ \frac{2}{p^2 N}\CT \left( \Sigma_{m,R}^{\text{NP},1}+ \CT\left( - \widehat{Z}_\lambda^{(1)} + \widehat{Z}_m^{(1)} - Z_\psi^{(1)} \right) \right).
\label{eq_twopoint_C2}
\end{multline}
Due to the factorization (\ref{ft2}), the first term in the second line corresponds to the diagrams in \figref{Nreducible-fig-3}--\ref{Nreducible-fig-4}, while the second term in the second line corresponds to the diagrams in \figref{Nreducible-fig-1}--\ref{Nreducible-fig-2} (in both cases, up to this order in $1/N$). Only the first line remains to be accounted for in \eqref{eq_twopoint_C2}, which has to be given by the diagram in \figref{4quarkL-fig} plus irreducible diagrams of order $1/N$, i.e. those that are not included in \figref{Nreducible-fig}. Including the renormalization constants in \eqref{V-condensate-renormalization}, we obtain
\be
\label{4qself}
 {1\over N}\Sigma_{p,R}^{{\rm NP}}= {\CF -\CT^2 \over p^2} + {\CF \over p^2 N }  \left( \widehat Z_\lambda^{(1)} +\widehat Z_V^{(1)}- 2\left( \widehat Z_\lambda^{(1)} -\widehat Z_m ^{(1)}\right) \right) -\ri \slashed p \times \text{irreducible}+ \CO\bigl(N^{-2}\bigr).
 \ee
The first term in (\ref{4qself}) can be written 
more explicitly by reexpressing the condensates in terms of $\Lambda$, through (\ref{cond-lambda}) and (\ref{4quark-L}). We find 
\be
\label{FT}
\CF-\CT^2 = { \Lambda^2  \over N} \left( d_1-2 c_1 + \log(\lambda) -1 + {\beta^{(1)}_\lambda(\lambda) \over \lambda^2} + 2 \int_0^\lambda {\chi(u)- \chi_1 u \over u^2} \rd u  \right) + \CO\bigl(N^{-2}\bigr), 
\ee
where the coefficients $d_1$, $c_1$ were introduced in (\ref{cNas}), (\ref{dNexp}). Based on the general arguments in \cite{david2,david3}, we expect $d_1$, the $1/N$ correction to the four-quark condensate, to be ambiguous. We will see that this 
is indeed the case. 

Let us note that the second term in the r.h.s. of (\ref{4qself}) should cancel the divergences obtained from these diagrams. Let us find a more convenient form for this 
combination of renormalization constants.  
We first derive from (\ref{zlambda}) the integral form
\be
 \widehat Z_\lambda^{(1)}= \int_0^\lambda {\epsilon \beta^{(1)}_\lambda (u)  \over u^2 (u+ \epsilon)^2} \rd u. 
\ee
After integration by parts we obtain the convenient expression
\be
\widehat Z_\lambda^{(1)}+   \widehat  Z_{V}^{(1)}=  \int_0^\lambda {\rho(u) \over u(u+\epsilon)} \rd u, 
\ee
where 
\be
\rho(\lambda)= -\lambda^2  {\rd \over \rd \lambda}\left( {\beta^{(1)}_\lambda(\lambda) \over \lambda^2} \right). 
\ee
The quantity $\widehat Z_\lambda^{(1)} -\widehat Z_m ^{(1)}$ is known from the calculation of the two-quark condensate (\ref{zz-2}). Therefore, 
the singular part in the irreducible diagrams that contribute to the four-quark condensate determines the beta function at NLO in $1/N$. 
As we will see, this calculation is simpler than the one usually adopted for the calculation of the beta function at this order in \cite{pmp, ab}. 
By using the known value (\ref{beta1}), we have 
\be
\rho(\lambda)= - {\lambda^2 \Gamma(2+ \lambda) \over  (2+\lambda)  \Gamma \left(1-\frac{\lambda
   }{2}\right) \Gamma^3 \left( 1 + \frac{\lambda }{2}\right)} =-\lambda \chi(\lambda).  
   \ee
We can now write
\be
\label{div-quartic}
\widehat  Z_\lambda^{(1)} +\widehat Z_V^{(1)}- 2\left(\widehat  Z_\lambda^{(1)} -\widehat Z_m ^{(1)}\right)=  \int_0^\lambda  {\upsilon (u) \over u(u+ \epsilon)}  \rd u,
\ee
where
\be
\label{ups-f}
\upsilon(u)=- {u \Gamma(2+ u) \over   \Gamma^3 \left(1+\frac{u }{2}\right) \Gamma \left(1-\frac{u
   }{2}\right)}. 
   \ee

\begin{figure}[!ht]
\centering
\includegraphics[height=3cm,trim={0 1.3cm 0 2cm},clip]{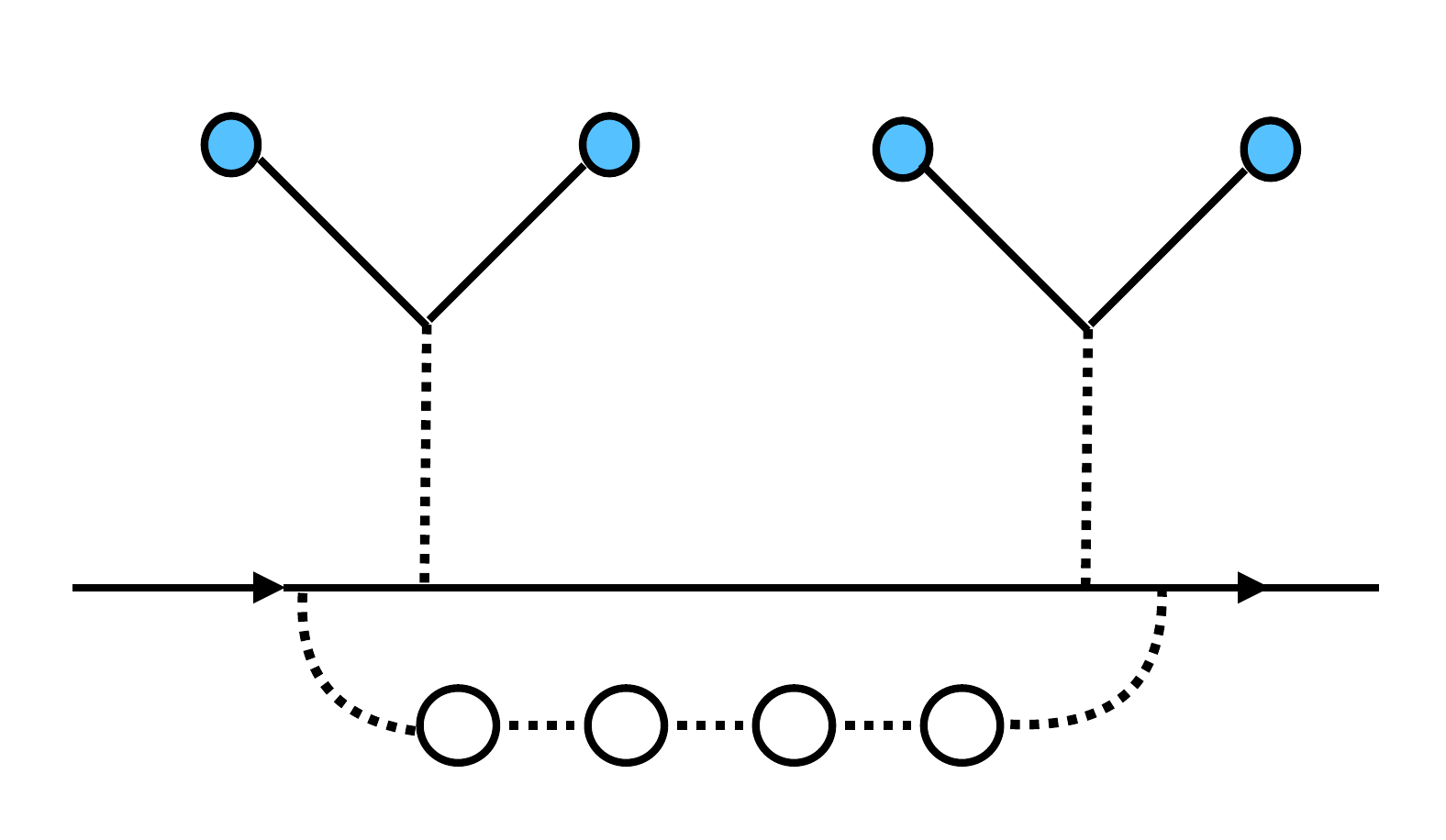}
\caption{Irreducible diagram that contributes to the four-quark condensate.}
\label{4quark-main-fig}
\end{figure}

\begin{figure}[!ht]
\begin{equation*}
\raisebox{-1.9cm}{\includegraphics[height=4cm,trim={0 2cm 23cm 2cm},clip]{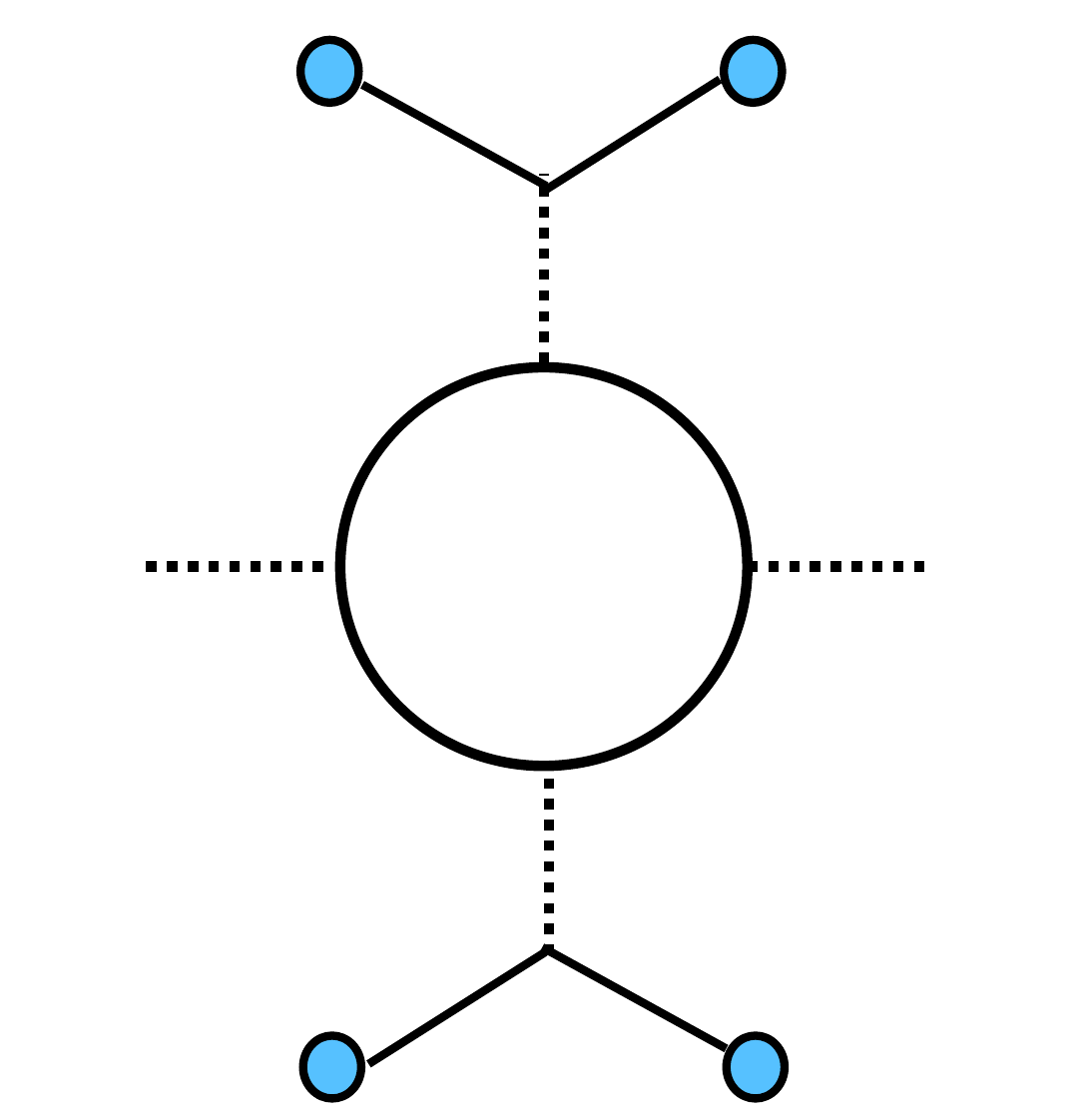}}
+ 2 \times
\raisebox{-1.9cm}{\includegraphics[height=4cm,trim={21cm 2cm 1.5cm 2cm},clip]{4quark-balls-fig.pdf}}
\end{equation*}
\caption{Two possible ways of inserting a four-quark condensate in a fermion polarization loop. The second diagram needs a symmetry factor 2.}
\label{4quark-balls-fig}
\end{figure}

\begin{figure}[!ht]
\centering
\includegraphics[width=10cm,trim={0 2.5cm 0 1.2cm},clip]{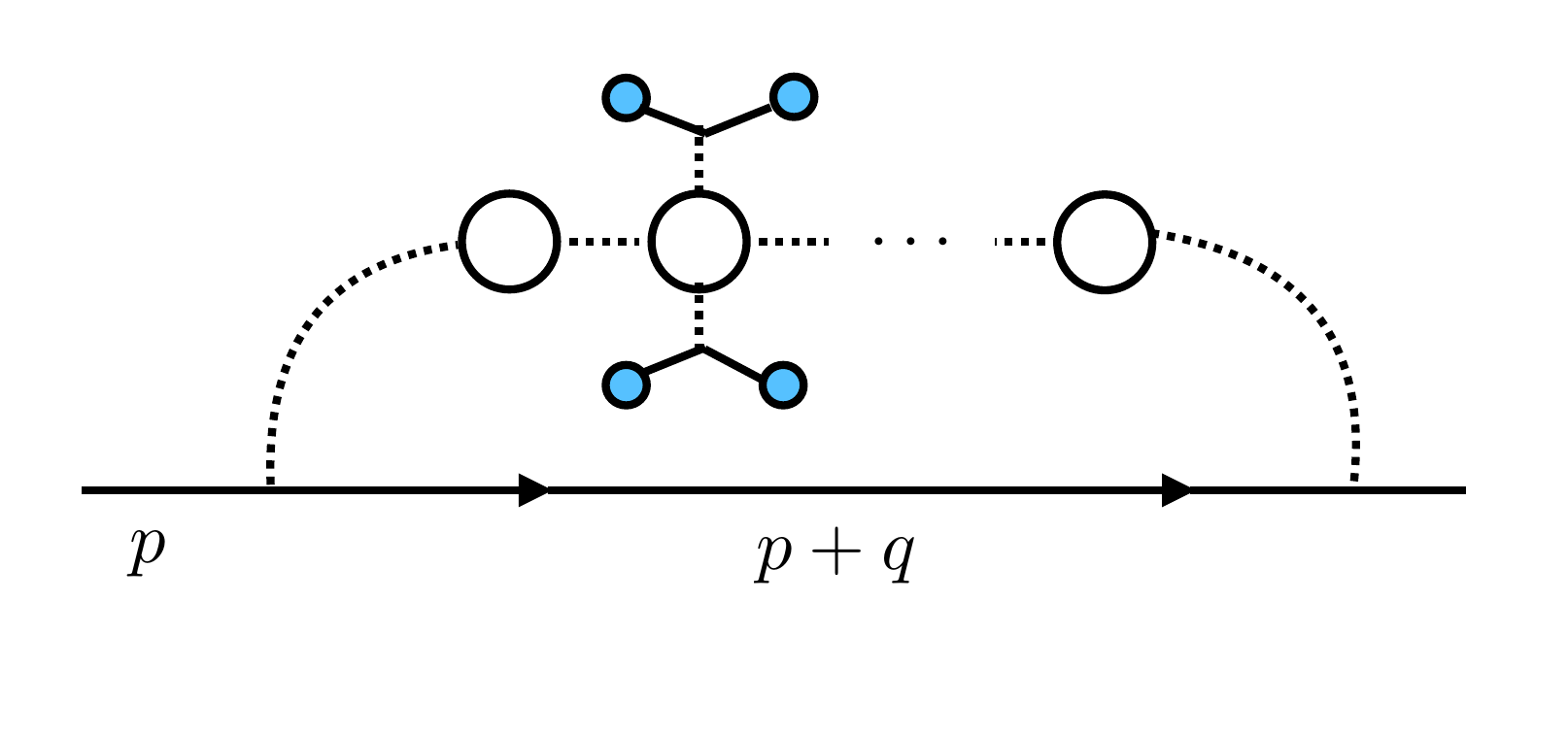}
\caption{A chain of bubbles with an insertion of one of the ``decorated" polarization loops.}
\label{chain-4quark-fig}
\end{figure}

\begin{figure}[!ht]
\centering
\includegraphics[width=10cm,trim={0 1.4cm 0 2.2cm},clip]{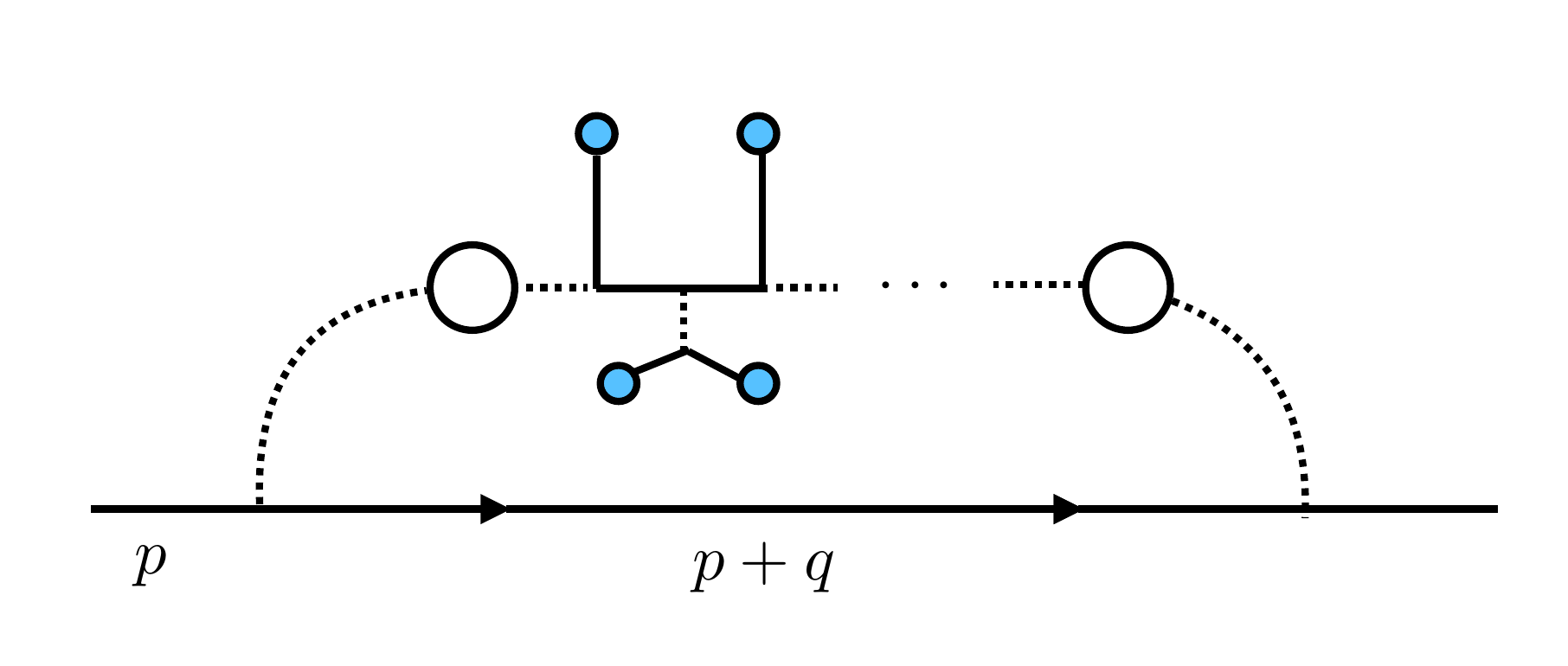}
\caption{A chain of bubbles with an insertion of a four-quark condensate in the propagator of the sigma field.}
\label{chain-4quark-2-fig}
\end{figure}

\begin{figure}[!ht]
\centering
\includegraphics[height=2.5cm,trim={0 1.4cm 0 1.3cm},clip]{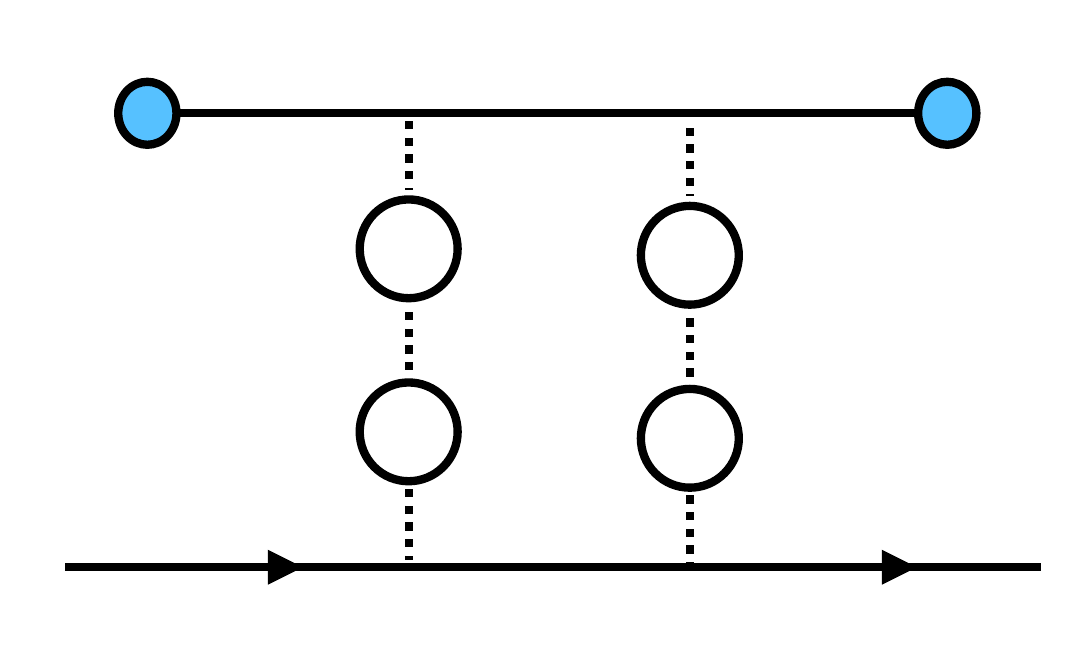}
\caption{When we expand the two-quark condensate around the same point, this diagram gives a contribution to the four-quark condensate.}
\label{hidden-4quark-fig}
\end{figure}

We have then to find the irreducible diagrams, to order $1/N$ and all loops. There are four types of diagrams that contribute: 

\begin{enumerate}

\item  The first type of diagram is obtained by putting 
a four-quark condensate in the middle of the internal propagator line of the bubble chain in \figref{chain-bubble-fig}. This gives the diagram shown in \figref{4quark-main-fig}. 

\item The second type of diagram is obtained by noticing that 
one can insert a four-quark condensate in a fermion polarization loop, in two different ways, to obtain a ``decorated" polarization loop, 
as shown in \figref{4quark-balls-fig}. This ``decorated" loop can then be inserted at any point inside a bubble chain, and leads to diagrams like the one in \figref{chain-4quark-fig}.     

\item The third type of diagram is obtained by noticing that 
one can insert a four-quark condensate in the sigma propagator, which can then be inserted in a bubble chain, as shown in \figref{chain-4quark-2-fig}. 

\item Finally, there are contributions coming from diagrams involving a two-quark condensate, in which the quark and antiquark are at different points, and one has to expand. 
The diagrams that contribute to this are the diagram in \figref{condensateN-fig-1}, and the diagram in \figref{hidden-4quark-fig}. Note that the latter gives a vanishing 
contribution to the two-quark condensate, but not to the four-quark condensate. 

\end{enumerate}

Let us now compute the contribution of the irreducible diagrams. It will be convenient to group them in appropriate ways.

We first consider the irreducible diagram in \figref{4quark-main-fig}, which we will combine with the diagram in \figref{condensateN-fig-1}. The 
contribution of \figref{4quark-main-fig} to $\Sigma_p^{\rm NP}$ is of the form 
\be
\label{fig8}
-{\pi \lambda_0   \over N^2} \frac{\langle   V \rangle_c}{p^2} \sum_{n \ge 1} \ri \CI_2(n) \lambda_0^n, 
\ee
where 
\be
\CI_2(n)= {p^2\over \slashed p} \pi^n \int {\rd^d q  \over  (2 \pi)^d} {\slashed p- \slashed q \over (p-q)^4} \left( \ri \Pi (q^2) \right)^{n-1}. 
\ee

Let us now consider the diagram in \figref{condensateN-fig-1}. In the previous section we calculated 
the first term in the expansion in the first line of (\ref{2-expand}), and we have 
to consider now the second term, which will produce a factor of $\langle K\rangle_c$. 
Its contribution to the diagram in position space involves an integral of the form
\be
\int \rd^d y_1 \rd^d y_2 (y_1- y_2)^\rho \re^{\ri (p-q) y_1 + \ri (q- k) y_2} \left(\Pi(q^2) \right)^n
\ee
where $p$ is the external momentum, $k$ and $q$ are internal momenta to be integrated over, 
and $n$ is the number of inserted bubbles. To calculate this integral, we write
\be
(y_1- y_2)^\rho \re^{\ri q(y_2-y_1)} =- {1 \over  \ri} {\partial \over \partial q_\rho} \re^{\ri q(y_2-y_1)} 
\ee
and we integrate by parts (see e.g. \cite{pascual-tarrach, elias} for similar calculations). After taking everything into account, we find that this 
diagram gives a contribution to $\Sigma_p^{\rm NP}$ of the form 
\be
-{\pi \lambda_0 \over dN^2}\langle K \rangle_c \sum_{n \ge 1}  {\partial (\ri \pi \Pi(p^2))^n  \over  \partial  p^2} \lambda_0^n, 
\ee
which will combine with (\ref{fig8}) into 
\be
\label{comb-in}
{\pi \lambda_0 \over N^2} \frac{\langle V \rangle_c}{p^2} \sum_{n \ge 1} \left( {p^2\over d}  {\partial (\ri \pi \Pi(p^2))^n  \over  \partial  p^2} - \ri \CI_2(n)  \right)  \lambda_0^n, 
\ee
after using (\ref{KV}). The general formula (\ref{gen-loop}) gives 
\be
\CI_2(n) = \ri \left(-\frac{p^2}{4\pi}\right)^{-n\epsilon/2} \frac{\Gamma \left(-\frac{\epsilon }{2}\right) \Gamma \left(1+\frac{n \epsilon }{2}\right) \Gamma \left(1-\frac{n \epsilon }{2}\right)}{\Gamma \left(\frac{(n-1) \epsilon}{2} \right) \Gamma \left(1-\frac{(n+1) \epsilon}{2} \right)} \left[ -\frac{1}{2}  \frac{\Gamma\left(1+\frac{\epsilon}{2}\right)\Gamma\left(1-\frac{\epsilon}{2}\right) \Gamma\left(-\frac{\epsilon}{2}\right) }{\Gamma(1-\epsilon)} \right]^{n-1}.
\ee
On the other hand, 
\begin{equation}
\begin{multlined}[b][0.8\textwidth]
{p^2\over d}  {\partial (\ri \pi \Pi(p^2))^n  \over  \partial  p^2} = -{1\over 4} \left(-\frac{p^2}{4\pi}\right)^{-n\epsilon/2} \left(-{n\epsilon \over 2} \right) \frac{\Gamma\left(1+\frac{\epsilon}{2}\right)\Gamma\left(1-\frac{\epsilon}{2}\right)\Gamma\left(-\frac{\epsilon}{2}\right)}{\left(1-\frac{\epsilon}{2}\right)\Gamma(1-\epsilon)}\\
\times \left[ -\frac{1}{2}  \frac{\Gamma\left(1+\frac{\epsilon}{2}\right) \Gamma\left(1-\frac{\epsilon}{2}\right) \Gamma\left(-\frac{\epsilon}{2}\right)}{\Gamma(1-\epsilon)} \right]^{n-1}.
\end{multlined}
\end{equation}
Combining both, and using the technique of Appendix \ref{pmp-trick}, we find that the sum in (\ref{comb-in}) is governed by the structure function
\begin{multline}
M(x,y)=- {1 \over 4}  \left(-\frac{p^2}{4 \pi \nu^2} \right)^{-y/2} \left[\frac{\Gamma \left(1+\frac{x}{2}\right) \Gamma^2 \left(1-\frac{x}{2}\right)}{\Gamma (1-x)}\right]^{y/x-1} \\
 \times y \left[ \frac{\Gamma \left(-\frac{x}{2}\right) \Gamma \left(1+\frac{y}{2}\right) \Gamma \left(1-\frac{y}{2}\right)}{\Gamma \left(\frac{y-x}{2}\right) \Gamma \left(1-\frac{y+x}{2}\right)}-{y \over 2} \frac{\Gamma \left(1+\frac{x}{2}\right) \Gamma \left(1-\frac{x}{2}\right)\Gamma\left(-\frac{x}{2}\right)}{ \left(1- {x \over 2} \right)\Gamma (1-x)} \right].
\end{multline}
Note also 
that $M_0(x)=0$. This means that the sum \eqref{comb-in} can be made finite simply by renormalizing the coupling constant. We also find 
\be
M(0,y)= \frac{y^2}{8}  \left(-1 + 2\gamma_E + \psi\left( 1- \frac{y}{2} \right) + \psi\left(\frac{y}{2}\right)
   \right). 
   \ee
We conclude that the contribution of these two classes of diagrams to $\Sigma_p^{\rm NP}$ is 
\be
-{m^2 \over p^2}  \Biggl\{ {\lambda \over 4}+{\lambda^2 \over 8} +\sum_{k \ge 1} {(2k+1)! \over 2^{2k+2}} \zeta(2k+1)\lambda^{2k+2} \Biggr\}.  
\label{eq_final_fig11}
\ee

We will now consider the combination of the diagrams in \figref{chain-4quark-fig} and \figref{chain-4quark-2-fig}. First of all, 
we calculate the amplitude associated to the ``decorated" loops in \figref{4quark-balls-fig}. We find
\be
\label{dl}
\Pi_{(\overline \psi \psi)^2}(p^2)=2 \int {\rd^d q \over (2 \pi)^d} \left[ {3 \over q^2 (q-p)^2}- {2 q\cdot p \over q^4 (q-p)^2} \right]=- {2 \over p^2} (1-\epsilon) \Pi(p^2).
\ee
Here we have included both the sign $-1$ due to the fermionic loop and the $-1=(-\ri)^2$ coming from the two $\sigma$ propagators. The symmetry factor of the diagram in \figref{chain-4quark-fig}, in which there are $n-2$ conventional polarization loops 
and one decorated loop, is $n-1$. The diagram in \figref{chain-4quark-2-fig}, in which there are $n-1$ polarization bubbles, has a symmetry factor $2n$. The contribution of these two classes of diagrams to the self-energy is 
\be
\label{dec-chain}
\frac{\pi \lambda_0}{N^2} \frac{\langle V \rangle_c}{p^2} \sum_{n \ge 1} \left( -2 n \ri \CI_3 (n) - (n-1) \ri \CI_4(n) \right) \lambda_0^n, 
\ee
where 
\be
\CI_3(n)= \frac{p^2}{\slashed{p}} \pi^n  \int {\rd^d q \over (2 \pi)^d} {\slashed p -\slashed q \over q^2 (p-q)^2} \left(\ri \Pi(q^2)  \right)^{n-1}
\ee
is the integral associated to the diagram in \figref{chain-4quark-2-fig}, while 
\be
  \CI_4(n)= \frac{p^2}{\slashed{p}} \pi^n \int {\rd^d q \over (2 \pi)^d} {\slashed p -\slashed q \over (p-q)^2} \left( \ri \Pi(q^2)  \right)^{n-2}\left( \ri \Pi_{(\overline \psi \psi)^2} (q^2)  \right)  
  \ee
  is the integral associated to the diagram in \figref{chain-4quark-fig}. In view of (\ref{dl}), the two integrals are related as
\be
\CI_4(n) = -2(1-\epsilon) \CI_3(n).
\ee
The integral $\CI_3 (n)$ can be computed with the expression (\ref{gen-loop}), and we find that the sum in (\ref{dec-chain}) is governed 
by the structure function 
\be
R(x, y)= \left(-\frac{p^2}{4 \pi \nu^2} \right)^{-y/2} (1+y-x) { \Gamma \left(1-{x \over 2} \right) \Gamma \left(1-{y \over 2} \right) \Gamma \left(1+{y \over 2} \right)  \over \Gamma \left(1+{y-x \over 2} \right) \Gamma \left(1-{y+x \over 2} \right)} 
\left[ {\Gamma \left(1+{x \over 2} \right) \Gamma^2 \left(1-{x \over 2} \right) \over \Gamma(1-x)} \right]^{y/x-1}. 
\ee
We note that 
\be
\label{rx0}
R_0(x)= { \Gamma \left(2-x \right) \over \Gamma^3 \left(1-{x\over 2} \right) \Gamma \left(1+{x \over 2} \right) }= {\upsilon (-x) \over x},  
\ee
where $\upsilon (x)$ was introduced in (\ref{ups-f}). We also have 
\be
\label{r0y}
R(0,y)= 1+y
\ee
for $\mu^2= - p^2$. Let us calculate the finite and divergent parts due to (\ref{rx0}). We have 
\be
\left[ R_0(\epsilon) \log\left( 1+ {\lambda \over \epsilon} \right) \right]_{\rm div}= \int_0^\lambda {R_0(-u) \over u+ \epsilon} \rd u = 
-\int_0^\lambda {\upsilon (u) \over u(u+\epsilon)} \rd u. 
\ee
This will cancel precisely the divergent part in (\ref{div-quartic}). Therefore, the diagrams of \figref{chain-4quark-fig} and \figref{chain-4quark-2-fig} are the 
relevant ones to compute the anomalous dimension of the operator $(\overline \bpsi \bpsi)^2$ and, therefore, of the beta function at NLO. 
A related calculation of this anomalous dimension 
in \cite{giombi} uses these diagrams in disguise.

The finite part due to $R_0(x)$ is given by
\be
-\int_0^\lambda{R_0(-u)-1 \over u} \rd u= 1-{\beta_\lambda^{(1)} (\lambda) \over \lambda^2} -2 \int_0^{\lambda} {\chi(u)- \chi_1 u \over u^2} \rd u,  
\ee
and it cancels with part of the expression in (\ref{FT}), leaving us with the following contribution to $\Sigma_{p, R}^{\rm NP}$:
\begin{equation}
\frac{m^2}{p^2} (d_1 -2 c_1 + \log(\lambda)).
\end{equation}
This result combines with the finite part coming from (\ref{r0y}), resulting in the contribution
\be
{m^2 \over p^2} (d_1 -2 c_1 + \log(\lambda) + \lambda).
\label{eq_final_fig12}
\ee

Let us finally consider the contribution of the family of diagrams in \figref{hidden-4quark-fig}. When the two quarks are at the same point, this diagram is proportional to $\tr (\gamma^\mu)$ and it vanishes. When we expand the two quark fields around the same point, we find 
a contribution to the four-quark condensate which involves an integral of the form
\be
 \int \rd^d y_1 \rd^d y_2  \, \re^{\ri (p-k-q) y_1 - \ri (r-k - q) y_2 } (y_1-y_2)^\rho {\ri \over \slashed k} {q_\rho \over q^2},  
\ee
where $p$ is the external momentum and $k, q,r$ are internal momenta to be integrated over. In the integration by parts one obtains
 \be
 {\partial \over \partial q_\rho} \left( {q_\rho \over q^2} \right)= {d-2 \over q^2}. 
 \ee
The contribution of the sum over all bubbles is 
\be
- \frac{\pi\lambda_0}{N^2} \frac{\langle V \rangle_c}{p^2}   \sum_{n \ge 1} \frac{d-2}{d} 2 n \ri \CI_3(n) \lambda_0^n,
\label{eq_fig13}
\ee
which can be calculated in terms of the structure function
\be
S(x, y)= -  \left(-\frac{p^2}{4 \pi \nu^2} \right)^{-y/2} \frac{y}{2-x} {\Gamma \left(1-{x \over 2} \right) \Gamma \left(1-{y \over 2} \right)\Gamma \left(1+{y \over 2} \right) \over \Gamma \left(1-{y+x \over 2} \right)\Gamma \left(1+{y-x \over 2} \right)} 
\left[ {\Gamma \left(1+{x \over 2} \right) \Gamma^2 \left(1-{x \over 2} \right) \over \Gamma(1-x)} \right]^{y/x-1}.
\ee
We find $S_0(x)=0$, and 
\be
S(0,y)= -{y \over 2}.
\ee
Including the factors in front of the sum in \eqref{eq_fig13}, we obtain a contribution to $\Sigma_{p, R}^{\rm NP}$ of the form  
\be
- {m^2 \over p^2} { \lambda \over 2}.
\label{eq_final_fig13}
\ee

The results in \eqref{eq_final_fig11}, \eqref{eq_final_fig12} and \eqref{eq_final_fig13} combine into
\be
\Sigma_{p, R}^{\rm NP} = -{m^2  \over p^2}  \biggl\{ 2 c_1-d_1 -\log(\lambda)  -{\lambda \over 4}+ {\lambda^2 \over 8} +\sum_{k \ge 1} {(2k+1)! \over 2^{2k+2}}  \zeta(2k+1)\lambda^{2k+2} \biggr\}+ \CO\bigl(N^{-1}\bigr). 
\ee
This agrees precisely with the result in (\ref{ts-sigma}), involving the series $\Phi_2(\lambda)$ of (\ref{ts1}), except for the constant terms, which 
depend on the values of the condensates. Moreover, by comparing the two results we can read off the value of $d_1$, giving the $1/N$ correction 
to the four-quark condensate
\be
\label{d1-value}
d_1= 1-\gamma_E + 2 \log(2) \mp {\ri \pi \over 2}. 
\ee
As advertised, this is ambiguous due to the renormalon in the perturbative series, and the choice of sign in (\ref{d1-value}) should be correlated with a choice of resummation prescription for the series $\Phi_0(\lambda)$. 

\begin{figure}[!ht]
\centering
\includegraphics[width=12cm]{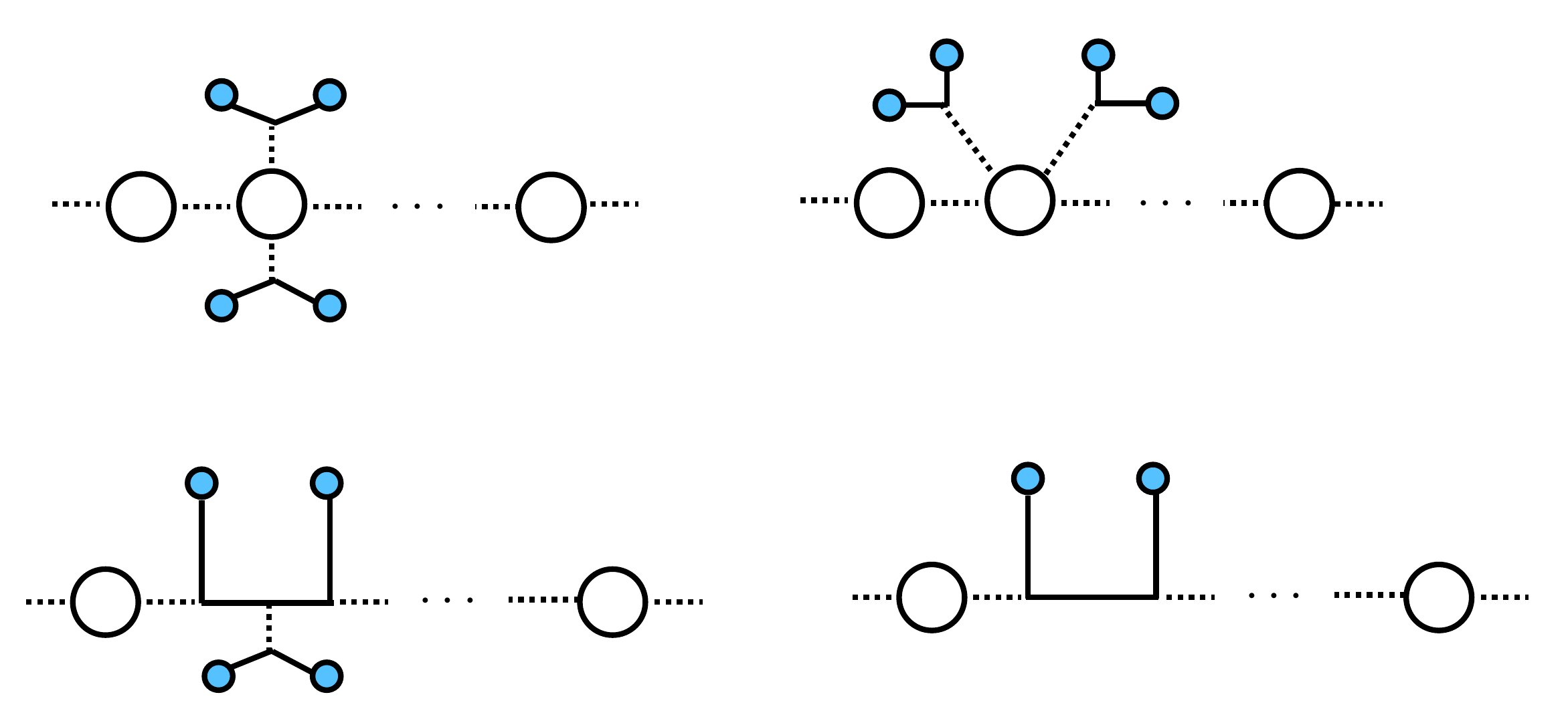}
\caption{The diagrams contributing to the four-quark condensate correction to the sigma propagator, at order $1/N$.}
\label{sigmap-fig}
\end{figure}
\noindent

A simple corollary of this calculation is a determination of the four-quark condensate contribution to the propagator of the sigma particle, at order $1/N$. Diagrammatically it is given by the sum of bubbles with condensate insertions shown in \figref{sigmap-fig}. By using the results (\ref{dec-chain}) and (\ref{eq_fig13}), one finds
 \be
 \label{lead-csigma}
 -{2\pi \ri\over N} {2 m_0^2 \over p^2} \left( { 1\over \log \left( -p^2 /m_0^2\right)} + {1 \over  \log^2  \left( -p^2 /m_0^2\right)}\right). 
 \ee
It is easy to check that this agrees with the result of expanding (\ref{D-ex}) at large $p^2 \gg m_0^2$. 

\sectiono{Conclusions and open problems}
\label{sec-conclusions}

The combination of the OPE with vacuum condensates, as developed in the SVZ sum rules, 
produces formal trans-series for QFT observables, and it provides a method to systematically calculate  exponentially small corrections to perturbative quantities. In this paper we have compared 
the result of an OPE calculation to an exact trans-series obtained in the $1/N$ expansion, and we have 
found complete agreement between the two, up to the unknown values of the condensates. 

In our calculation we have used the ``practical" version of the OPE. As we mentioned in the Introduction, 
it has been suggested that one should use a more complicated version, based on the introduction 
of an additional momentum scale. In this ``Wilsonian" version, IR renormalons are absent, since the additional 
scale provides an explicit IR cutoff for the Feynman integrals, and the condensates are defined unambiguously. The 
prize to pay is that each series in the trans-series depends on this scale, and the dependence only drops out in the 
total result. Our study of the GN model seems to confirm that there 
is nothing wrong with the ``practical" version of the OPE, in agreement with the discussions in \cite{david3,beneke-braun2}. 
The exact results for the two-point function 
can be decoded in terms of a trans-series in which condensates are ambiguous, but this ambiguity is due to the 
well-known Stokes phenomenon and does not lead to any inconsistency. In addition, the series in the 't Hooft parameter 
appearing in this trans-series can be reproduced exactly, and rather non-trivially, with the ``practical" OPE.

A nice outcome of our calculation is the following. In the exact large $N$ answer, 
the trans-series emerges as a formal, algebraic object. The OPE calculation gives 
a concrete picture of this trans-series in terms of perturbation theory with 
condensates. In particular, the factorial 
divergence of the perturbative series appearing in the trans-series is the manifestation of a renormalon-like phenomenon, due to bubble chains attached to the condensates, as illustrated in e.g. \figref{condensateN-fig} or \figref{4quark-main-fig}. Note that our calculation indicates that the only sources of exponentially small corrections in the exact trans-series are 
condensates, and in particular no instanton effects have been found. This is expected since we are working in 
the $1/N$ expansion. There are also indications that large $N$ instantons are absent in this model \cite{avan-vega,dmss}. 

Although this is probably well-known to many practitioners, it is worth noting that the 
OPE calculation reconstructs the trans-series more efficiently than a resurgent/renormalon 
analysis of the perturbative series. In particular, the two-quark condensate 
leads to a non-ambiguous power correction which is completely invisible in a resurgent analysis. 
The Borel singularity of the perturbative series detects the presence of a power correction in $m^2/p^2$ due to the four-quark condensate, but since this singularity is essentially a simple pole it misses the full series $\Phi_2(\lambda)$. The 
OPE calculation is in contrast able to reproduce the series 
$\Phi_{1,2}(\lambda)$ at all loops. It was suggested in \cite{dmss} that part of the 
``blindness" of resurgent analysis in this situation might be due to the restriction to a given order in 
the $1/N$ expansion. It might happen that at finite $N$ one can have a better access to the four-quark condensate 
through the resurgent structure of the perturbative series, but the two-quark condensate series will remain undetected, 
since it transforms non-trivially under the $\IZ_2$ chiral symmetry and does not mix with the identity operator \cite{david2}. 
In the language of \cite{dmss}, the trans-series for the self-energy at order $1/N$ does not satisfy the strong version of the resurgence program, 
since the resurgent analysis of the perturbative series does not make it possible to reconstruct the full trans-series. It does satisfy 
however the weak version of the program, since the exact result can be obtained by the (lateral) Borel resummation of the trans-series. 

Our calculation can be extended in many ways. As we mentioned in the Introduction, the 
GN model turns out to be a simpler example to study than bosonic sigma models in 
two dimensions, since in the latter the fields are constrained and in addition there is an infinite 
number of operators with a fixed scaling dimension. It would be very interesting however 
to reproduce the large $N$ trans-series obtained in \cite{beneke-braun} 
(or the supersymmetric version studied more recently in \cite{sss}) by an OPE calculation 
with condensates similar to the one done here. The results of 
\cite{svz-pr,sss} on non-linear sigma models might be a good starting point for this calculation.
 
Although going beyond the $1/N$ expansion is analytically difficult, we note that two-point functions can be computed 
non-perturbatively in integrable models by using form factors. In the case of the 
non-linear sigma model, it has been checked numerically that the form factor 
calculation reproduces asymptotically the perturbative series \cite{balog, balog-nh}. 
It would be very interesting to see whether it is possible to detect as well condensate 
corrections to the perturbative result through form factors.

Finally, although the results of this paper vindicate the idea that the OPE with vacuum condensates leads to the  
correct exponentially small corrections to the perturbative series, it is still not clear how the 
power corrections found in \cite{mmr-an,mmr-theta,abbh2,bbv2,schepers,bbv3} can be reproduced by a first 
principles calculation. 
This remains in our view a sharp open problem for our understanding of non-perturbative effects in QFT. 

\section*{Acknowledgements}
We would like to thank Martin Beneke, Tomás Reis and Marco Serone for very useful discussions and a 
careful reading of the draft. The work of M.M. has been supported in part by the ERC-SyG project 
``Recursive and Exact New Quantum Theory" (ReNewQuantum), which 
received funding from the European Research Council (ERC) under the European 
Union's Horizon 2020 research and innovation program, 
grant agreement No. 810573. The work of R.M. is supported by the NKFIH K134946 Grant.

\appendix

\sectiono{Bubbles}
\subsection{The fermion polarization loop}
\label{pol-app}

The building block of a bubble chain is the fermion polarization loop, given by the following diagram: 
\be
\begin{tikzpicture}[baseline=- 0.5ex]
\begin{feynman}
\vertex (a1);
\vertex[right=2 of a1] (a2);
\draw[fermion] (a1) arc(180:0:1);
\draw[fermion] (a2) arc(0:-180:1);
\fill (a1) circle(0.1) node[right];
\fill (a2) circle(0.1) node[left];
\vertex[left= 0.75 of a1] (b1);
\vertex[left= 0.2 of a1] (b2);
\draw[->] (b1) node[above right] {$p$} -- (b2);
\vertex[right= 0.75 of a2] (c1);
\vertex[right= 0.2 of a2] (c2);
\draw[->] (c2)  -- (c1) node[above left] {$p$};
\end{feynman}
\end{tikzpicture}
\label{eq_polarization_loop}
\ee
We define the corresponding amplitude, with massless fermions, as  
\be
\label{pol-loop}
 \Pi (p^2)= \int {\rd^d q \over (2 \pi)^d} \tr \left[ {1\over \slashed q  ( \slashed p + \slashed q)} \right] =2 \int {\rd^d q \over (2 \pi)^d} {q\cdot p \over q^2( p+q)^2}, 
\ee
where we used dimensional regularization to drop a scale-less integral. This and other integrals appearing in this paper can be computed with the 
master formula
\begin{multline}
\int \frac{\rd^d q}{(2\pi)^d} \frac{q^{(\mu_1 \mu_2\dots \mu_n)}}{(q^2)^r [(p-q)^2]^s} = -\frac{\ri}{4\pi} \left(-\frac{p^2}{4\pi}\right)^{-\epsilon/2} \frac{1}{(p^2)^{r+s-1}} p^{(\mu_1 \mu_2\dots \mu_n)}\\
\times \frac{\Gamma( 1 + n - r - \epsilon/2)\Gamma(1-s-\epsilon/2)\Gamma(r+s-1+\epsilon/2)}{\Gamma(r)\Gamma(s)\Gamma(2-r-s+n-\epsilon)}, 
\label{gen-loop}
\end{multline}
where $q^{(\mu_1 \mu_2\dots \mu_n)}$ is the traceless symmetric tensor constructed from $q^{\mu_1} q^{\mu_2} \dots q^{\mu_n}$, see Appendix C in 
\cite{pascual-tarrach} for more details. For example,
\be
q^{(\mu)}= q^\mu, \qquad q^{(\mu_1 \mu_2)}= q^{\mu_1} q^{\mu_2}-{1\over d} g^{\mu_1 \mu_2} q^2, 
\ee
but we will only need to compute integrals with one index at most. Using (\ref{gen-loop}), we find
\be
\label{pi-massless}
 \Pi(q^2) = \frac{\ri}{2\pi} \left(-\frac{q^2}{4\pi}\right)^{-\epsilon/2} \frac{\Gamma\left(1+\frac{\epsilon}{2}\right)\Gamma\left(1-\frac{\epsilon}{2}\right)\Gamma\left(-\frac{\epsilon}{2}\right)}{\Gamma(1-\epsilon)}.
\end{equation}

Let us now briefly consider the massive case. The massive 
polarization loop is defined as
\be
\label{mpol-loop}
 \Pi_{m_0} (p^2)= \int {\rd^d q \over (2 \pi)^d} \tr \left[ {1\over (\slashed q -m_0) ( \slashed p + \slashed q-m_0)} \right]. 
\ee
Using standard Feynman integral techniques, one finds in $d=2-\epsilon$ dimensions 
the $\epsilon$ expansion (see e.g. \cite{kleinert} for additional details)
 \be
 \label{exp-pim}
 \Pi_{m_0}(p^2)= -{\ri \over \pi \epsilon} + {\ri\over 2 \pi} \log \left( {m_0^2 \over 4 \pi \re^{-\gamma_E}} \right) 
 +{\ri \over 2 \pi} \xi \log\left[ {\xi+1 \over \xi-1} \right] + \CO(\epsilon), 
 \ee
where $\xi$ was introduced in (\ref{xi-def}). The $\sigma$ propagator in momentum space is closely related to the massive 
polarization loop:
\be
\label{D-int-rep}
\Delta^{-1}(p;m_0)= \Pi_{m_0}(p^2)- {1\over m_0}\int {\rd^d q \over (2 \pi)^d}  \tr \left[  {1\over \slashed q-m_0} \right], 
\ee
and by using (\ref{exp-pim}) one finds the expression (\ref{D-ex}). 

\subsection{Summing over bubbles}
\label{pmp-trick}

In \cite{pmp, pm}, a powerful and simple technique was introduced to obtain the renormalized 
perturbative series associated to a chain of bubbles. We now summarize its main ingredients. 

Let us consider a generic sum of corrections in the bare 't Hooft coupling $\lambda_0$, of order $1/N$:
\be
A =  \sum_{n \ge n_0} a_n (p^2)   \lambda_0^n, 
\ee
where $n_0 \ge 1$. The choice of $\lambda_0$ is such that, at leading order in the $1/N$ expansion, the renormalization function 
is given by (\ref{zl}). Since the quantity we are considering is already of order $1/N$, we only have to use the 
leading term $Z^{(0)}_\lambda$. Then the renormalized sum is of the form 
\be
\label{ren-sum}
A = \sum_{n \ge n_0} (\nu^2)^{n\epsilon/2} a_n (p^2)  \left(Z^{(0)}_\lambda \right)^n \lambda^n. 
\ee
Let us assume that one can find a ``structure function" $F(x,y)$, which is analytic in both arguments at $x=0$, $y=0$, and satisfying
\begin{equation}
(\nu^2)^{n\epsilon/2} a_n (p^2) = \frac{F(\epsilon,n\epsilon)}{n\epsilon^n}.
\label{eq_structure_function}
\end{equation}
Then the renormalized sum is of the form
\be
A = \sum_{n \ge n_0} f_n \lambda^n \left(1+{ \lambda \over\epsilon} \right)^{-n}, 
\ee
where we have abbreviated 
\be
f_n={F(\epsilon, n \epsilon) \over n \epsilon^n}. 
\ee
We now expand the factor $(1+\lambda/\epsilon)^{-n}$ in powers of $\lambda$, using the binomial theorem, 
and we get
\be
\ba
A &= \sum_{n \ge n_0} f_n \lambda^n \sum_{s \ge 0} {n+s-1 \choose s} (-1)^s \left( {\lambda \over \epsilon} \right)^s= \sum_{m \ge n_0} \lambda^m \sum_{s=0}^{m-1} {m-1 \choose s}{ (-1)^s \over \epsilon^s} f_{m-s} \\
&=\sum_{m \ge n_0} \left( {\lambda \over \epsilon} \right)^m \sum_{s=0}^{m-1} {m-1 \choose s} (-1)^s \sum_{j \ge 0} F_j (\epsilon) 
(m-s)^{j-1} \epsilon^j \\
&= \sum_{m \ge n_0} \lambda^m \sum_{j \ge 0} {F_j (\epsilon) \over \epsilon^{m-j}} \sum_{s=0}^{m-1} {m-1 \choose s} (-1)^s (m-s)^{j-1}, 
\ea
\ee
where we set $m=n+s$ and we performed the following expansion of the structure function:
\be
F(x,y)= \sum_{j \ge 0} F_j(x) y^j. 
\ee
We now use that
\be
\sum_{s=0}^{m-1} {m-1 \choose s} (-1)^s (m-s)^{j-1}= \begin{cases} {(-1)^{m-1} \over m} , & \text{if $j=0$}, \\
0, & \text{if $1\le j\le m-1$}, \\
(m-1)!, & \text{if $j=m$},
\end{cases}
\ee
to write the renormalized sum as 
\be
\label{rs-F}
A = F_0(\epsilon) \sum_{m \ge n_0} {(-1)^{m-1} \over m}  \left( {\lambda \over \epsilon} \right)^m  + \sum_{m \ge n_0} (m-1)! F_m (\epsilon) \lambda^m+ \CO(\epsilon).
\ee
The first series can also be expressed as
\be
F_0(\epsilon) \sum_{m \ge n_0} {(-1)^{m-1} \over m}  \left( {\lambda \over \epsilon} \right)^m = F_0(\epsilon)\log\left(1+{\lambda \over \epsilon} \right)-F_0(\epsilon) \sum_{m=1}^{n_0-1}{(-1)^{m-1}  \over m}  \left( {\lambda \over \epsilon} \right)^m.
\label{rs-F-2}
\ee

We now consider the behavior as $\epsilon \to 0$. There is a divergent part encoded in \eqref{rs-F-2}. We note the useful formula
\be
\label{F0div}
\left[ F_0(\epsilon) \log\left(1+ {\lambda \over \epsilon} \right) \right]_{\rm div}=  \int_0^\lambda {F_0(-u) \over u+ \epsilon} \rd u, 
\ee
and we can write the total divergent part in \eqref{rs-F-2} as
\be
\label{div-part}
[A]_\text{div} = \int_0^\lambda {F_0(-u) \over u+ \epsilon} \rd u - \sum_{m=1}^{n_0-1} {(-1)^{m-1} \over m} \lambda^m \sum_{k=0}^{m-1} F_{0,k} \epsilon^{k-m}, 
  \ee
where the coefficients $F_{0,k}$ are defined by 
\be
F_0(\epsilon)=  \sum_{k \ge 0} F_{0,k} \epsilon^k. 
\ee

The finite part of \eqref{rs-F} arises from the second sum plus the terms of order $\epsilon^0$ in the first sum:
\be
[A]_\text{finite} = \sum_{m \ge n_0} {(-1)^{m-1} \over m} F_{0,m} \lambda^m + \sum_{m \ge n_0} (m-1)! F_m (\epsilon) \lambda^m.
\ee
It will be convenient to sum the first series into an integral, yielding
\be
\label{finite-part}
[A]_\text{finite} = -\int_0^\lambda {F_0(-u) -F_{0,0} \over u} \rd u- \sum_{m=1}^{ n_0-1} {(-1)^{m-1} \over m} F_{0,m} \lambda^m+ \sum_{m \ge n_0} (m-1)! F_m (0) \lambda^m. 
\ee
The last sum in the above expression has the form of an inverse Borel transform.

As we have seen, the structure function \eqref{eq_structure_function} is the relevant object in diagrammatic computations, as it contains all the necessary information to extract both the divergent part \eqref{div-part} and the finite part \eqref{finite-part}.

Another useful result for the calculation of renormalization functions is the following. Let $f(\epsilon)$ be analytic at $\epsilon=0$. Then \cite{cr-dr}
\be
\label{use-id}
(\lambda+ \epsilon) {\partial \over \partial \lambda} \left[ \log\left( 1+ {\lambda \over \epsilon} \right) f(\epsilon) \right]_{\rm div} = f(-\lambda).  
\ee
This follows from a direct calculation: 
\be\ba
(\lambda+ \epsilon) {\partial \over \partial \lambda} \left[ \log\left( 1+ {\lambda \over \epsilon} \right) f(\epsilon) \right]_{\rm div} &= 
\sum_{m \ge 1} (-1)^{m-1} (\lambda^m + \epsilon \lambda^{m-1}) \sum_{k =0}^{m-1} f_k  \epsilon^{k-m}\\
&=\sum_{m \ge 0} (-1)^m f_m \lambda^m. 
\ea
\ee

\bibliographystyle{JHEP}

\linespread{0.6}
\bibliography{biblio-qft}

\providecommand{\href}[2]{#2}\begingroup\raggedright\begin{thebibliography}{10}

\bibitem{beneke-braun}
M.~Beneke, V.~M. Braun and N.~Kivel, \emph{{The Operator product expansion,
  nonperturbative couplings and the Landau pole: Lessons from the O(N) sigma
  model}}, \href{http://dx.doi.org/10.1016/S0370-2693(98)01339-2}{\emph{Phys.
  Lett.} {\bf B443} (1998) 308--316},
  [\href{http://arxiv.org/abs/hep-ph/9809287}{{\tt hep-ph/9809287}}].

\bibitem{david2}
F.~David, \emph{{On the Ambiguity of Composite Operators, IR Renormalons and
  the Status of the Operator Product Expansion}},
  \href{http://dx.doi.org/10.1016/0550-3213(84)90235-9}{\emph{Nucl. Phys.} {\bf
  B234} (1984) 237--251}.

\bibitem{david3}
F.~David, \emph{{The Operator Product Expansion and Renormalons: A Comment}},
  \href{http://dx.doi.org/10.1016/0550-3213(86)90279-8}{\emph{Nucl. Phys.} {\bf
  B263} (1986) 637--648}.

\bibitem{mmr-an}
M.~Mari\~no, R.~Miravitllas and T.~Reis, \emph{{New renormalons from analytic
  trans-series}}, \href{http://dx.doi.org/10.1007/JHEP08(2022)279}{\emph{JHEP}
  {\bf 08} (2022) 279}, [\href{http://arxiv.org/abs/2111.11951}{{\tt
  2111.11951}}].

\bibitem{mmr-theta}
M.~Mari\~no, R.~Miravitllas and T.~Reis, \emph{{Instantons, renormalons and the
  theta angle in integrable sigma models}},
  \href{http://dx.doi.org/10.21468/SciPostPhys.15.5.184}{\emph{SciPost Phys.}
  {\bf 15} (2023) 184}, [\href{http://arxiv.org/abs/2205.04495}{{\tt
  2205.04495}}].

\bibitem{abbh2}
M.~C. Abbott, Z.~Bajnok, J.~Balog, A.~Heged\'{u}s and S.~Sadeghian,
  \emph{{Resurgence in the $O(4)$ sigma model}},
  \href{http://dx.doi.org/10.1007/JHEP05(2021)253}{\emph{JHEP} {\bf 05} (2021)
  253}, [\href{http://arxiv.org/abs/2011.12254}{{\tt 2011.12254}}].

\bibitem{bbv2}
Z.~Bajnok, J.~Balog and I.~Vona, \emph{{The full analytic trans-series in
  integrable field theories}},
  \href{http://dx.doi.org/10.1016/j.physletb.2023.138075}{\emph{Phys. Lett. B}
  {\bf 844} (2023) 138075}, [\href{http://arxiv.org/abs/2212.09416}{{\tt
  2212.09416}}].

\bibitem{schepers}
L.~Schepers and D.~C. Thompson, \emph{{Asymptotics in an asymptotic CFT}},
  \href{http://dx.doi.org/10.1007/JHEP04(2023)112}{\emph{JHEP} {\bf 04} (2023)
  112}, [\href{http://arxiv.org/abs/2301.11803}{{\tt 2301.11803}}].

\bibitem{bbv3}
Z.~Bajnok, J.~Balog and I.~Vona, \emph{{Wiener-Hopf solution of the free energy
  TBA problem and instanton sectors in the O(3) sigma model}},
  \href{http://dx.doi.org/10.1007/JHEP11(2024)093}{\emph{JHEP} {\bf 11} (2024)
  093}, [\href{http://arxiv.org/abs/2404.07621}{{\tt 2404.07621}}].

\bibitem{stingl}
M.~Stingl, \emph{{Field theory amplitudes as resurgent functions}},
  \href{http://arxiv.org/abs/hep-ph/0207349}{{\tt hep-ph/0207349}}.

\bibitem{shifman-renormalons}
M.~Shifman, \emph{{New and Old about Renormalons: in Memoriam Kolya Uraltsev}},
  \href{http://dx.doi.org/10.1142/S0217751X15430010}{\emph{Int. J. Mod. Phys.
  A} {\bf 30} (2015) 1543001}, [\href{http://arxiv.org/abs/1310.1966}{{\tt
  1310.1966}}].

\bibitem{politzer}
H.~D. Politzer, \emph{{Effective Quark Masses in the Chiral Limit}},
  \href{http://dx.doi.org/10.1016/0550-3213(76)90405-3}{\emph{Nucl. Phys. B}
  {\bf 117} (1976) 397--406}.

\bibitem{svz}
M.~A. Shifman, A.~I. Vainshtein and V.~I. Zakharov, \emph{{QCD and Resonance
  Physics. Theoretical Foundations}},
  \href{http://dx.doi.org/10.1016/0550-3213(79)90022-1}{\emph{Nucl. Phys. B}
  {\bf 147} (1979) 385--447}.

\bibitem{svz2}
M.~A. Shifman, A.~I. Vainshtein and V.~I. Zakharov, \emph{{QCD and Resonance
  Physics: Applications}},
  \href{http://dx.doi.org/10.1016/0550-3213(79)90023-3}{\emph{Nucl. Phys. B}
  {\bf 147} (1979) 448--518}.

\bibitem{svz-ope}
V.~A. Novikov, M.~A. Shifman, A.~I. Vainshtein and V.~I. Zakharov,
  \emph{{Wilson's Operator Expansion: Can It Fail?}},
  \href{http://dx.doi.org/10.1016/0550-3213(85)90087-2}{\emph{Nucl. Phys. B}
  {\bf 249} (1985) 445--471}.

\bibitem{svz-pr}
V.~A. Novikov, M.~A. Shifman, A.~I. Vainshtein and V.~I. Zakharov,
  \emph{{Two-Dimensional Sigma Models: Modeling Nonperturbative Effects of
  Quantum Chromodynamics}},
  \href{http://dx.doi.org/10.1016/0370-1573(84)90021-8}{\emph{Phys. Rept.} {\bf
  116} (1984) 103}.

\bibitem{sss}
D.~Schubring, C.-H. Sheu and M.~Shifman, \emph{{Treating divergent perturbation
  theory: Lessons from exactly solvable 2D models at large N}},
  \href{http://dx.doi.org/10.1103/PhysRevD.104.085016}{\emph{Phys. Rev. D} {\bf
  104} (2021) 085016}, [\href{http://arxiv.org/abs/2107.11017}{{\tt
  2107.11017}}].

\bibitem{gross-neveu}
D.~J. Gross and A.~Neveu, \emph{{Dynamical Symmetry Breaking in Asymptotically
  Free Field Theories}},
  \href{http://dx.doi.org/10.1103/PhysRevD.10.3235}{\emph{Phys. Rev.} {\bf D10}
  (1974) 3235}.

\bibitem{cr-dr}
M.~Campostrini and P.~Rossi, \emph{{Dimensional regularization in the 1/N
  expansion}}, \href{http://dx.doi.org/10.1142/S0217751X92001459}{\emph{Int. J.
  Mod. Phys. A} {\bf 7} (1992) 3265--3290}.

\bibitem{pmp}
A.~Palanques-Mestre and P.~Pascual, \emph{{The 1/$N_f$ Expansion of the
  $\gamma$ and beta Functions in QED}},
  \href{http://dx.doi.org/10.1007/BF01212398}{\emph{Commun. Math. Phys.} {\bf
  95} (1984) 277}.

\bibitem{ab}
T.~Alanne and S.~Blasi, \emph{{The $\beta$-function for Yukawa theory at large
  $N_f$}}, \href{http://dx.doi.org/10.1007/JHEP08(2018)081}{\emph{JHEP} {\bf
  08} (2018) 081}, [\href{http://arxiv.org/abs/1806.06954}{{\tt 1806.06954}}].

\bibitem{landau}
K.~Langfeld, L.~von Smekal and H.~Reinhardt, \emph{{Landau pole screening in
  the Gross-Neveu model: 1/N versus operator product expansion}},
  \href{http://dx.doi.org/10.1016/0370-2693(95)01172-M}{\emph{Phys. Lett. B}
  {\bf 362} (1995) 128--133}.

\bibitem{pm}
A.~Palanques-Mestre, \emph{{Renormalons in QED}},
  \href{http://dx.doi.org/10.1007/BF01552504}{\emph{Z. Phys. C} {\bf 32} (1986)
  255}.

\bibitem{anselm}
A.~Anselm, \emph{{Field model with a nonvanishing renormalized charge}},
  {\emph{Zh. Eksp. Teor. Fiz.} {\bf 36} (1959) 363}.

\bibitem{shifman-anselm}
M.~Shifman, \emph{{Anselm's discovery of the Gross--Neveu model in 1958}},  in
  \emph{Under the spell of Landau: When theoretical physics was shaping
  destinies}, pp.~524--525.
\newblock World Scientific, 2013.
\newblock \href{http://dx.doi.org/10.1142/9789814436571_0050}{DOI}.

\bibitem{zz}
A.~B. Zamolodchikov and A.~B. Zamolodchikov, \emph{{Factorized S Matrices in
  Two-Dimensions as the Exact Solutions of Certain Relativistic Quantum Field
  Models}}, \href{http://dx.doi.org/10.1016/0003-4916(79)90391-9}{\emph{Annals
  Phys.} {\bf 120} (1979) 253--291}.

\bibitem{gracey-four}
J.~A. Gracey, T.~Luthe and Y.~Schroder, \emph{{Four loop renormalization of the
  Gross-Neveu model}},
  \href{http://dx.doi.org/10.1103/PhysRevD.94.125028}{\emph{Phys. Rev. D} {\bf
  94} (2016) 125028}, [\href{http://arxiv.org/abs/1609.05071}{{\tt
  1609.05071}}].

\bibitem{gracey-eta}
J.~A. Gracey, \emph{{Calculation of exponent eta to $O(1/N^2)$ in the $O(N)$
  Gross-Neveu model}},
  \href{http://dx.doi.org/10.1142/S0217751X91000241}{\emph{Int. J. Mod. Phys.
  A} {\bf 6} (1991) 395--408}.

\bibitem{gracey-mass}
J.~A. Gracey, \emph{{Anomalous mass dimension at $O(1/N^2)$ in the $O(N)$
  Gross-Neveu model}},
  \href{http://dx.doi.org/10.1016/0370-2693(92)91265-B}{\emph{Phys. Lett. B}
  {\bf 297} (1992) 293--297}.

\bibitem{collins}
J.~C. Collins, \emph{{Renormalization}}, vol.~26 of \emph{Cambridge Monographs
  on Mathematical Physics}.
\newblock Cambridge University Press, Cambridge, 1986,
  \href{http://dx.doi.org/10.1017/CBO9780511622656}{10.1017/CBO9780511622656}.

\bibitem{gran}
B.~Grinstein and L.~Randall, \emph{{The Renormalization of $G^{2}$}},
  \href{http://dx.doi.org/10.1016/0370-2693(89)90877-0}{\emph{Phys. Lett. B}
  {\bf 217} (1989) 335--340}.

\bibitem{robertson}
D.~G. Robertson, \emph{{Composite operator renormalization and the trace
  anomaly}}, \href{http://dx.doi.org/10.1016/0370-2693(91)91375-6}{\emph{Phys.
  Lett. B} {\bf 253} (1991) 143--148}.

\bibitem{bkkw-un}
B.~Berg, M.~Karowski, V.~Kurak and P.~Weisz, ``{Higher order $1/N$ calculations
  in the Gross-Neveu and nonlinear $\sigma$ models}.'' 1978.

\bibitem{bkkw}
B.~Berg, M.~Karowski, V.~Kurak and P.~Weisz, \emph{{Scattering Amplitudes of
  the {Gross-Neveu} and Nonlinear $\sigma$ Models in Higher Orders of the 1/$N$
  Expansion}},
  \href{http://dx.doi.org/10.1016/0370-2693(78)90916-4}{\emph{Phys. Lett. B}
  {\bf 76} (1978) 502--504}.

\bibitem{fnw2}
P.~Forgacs, F.~Niedermayer and P.~Weisz, \emph{{The Exact mass gap of the
  Gross-Neveu model. 2. The 1/N expansion}},
  \href{http://dx.doi.org/10.1016/0550-3213(91)90045-Y}{\emph{Nucl. Phys.} {\bf
  B367} (1991) 144--157}.

\bibitem{cr-review}
M.~Campostrini and P.~Rossi, \emph{{The 1/N expansion of two-dimensional spin
  models}}, \href{http://dx.doi.org/10.1007/BF02730034}{\emph{Riv. Nuovo Cim.}
  {\bf 16} (1993) 1--111}.

\bibitem{biscari}
P.~Biscari, M.~Campostrini and P.~Rossi, \emph{{Quantitative Picture of the
  Scaling Behavior of Lattice Nonlinear $\sigma$ Models From the 1/$N$
  Expansion}},
  \href{http://dx.doi.org/10.1016/0370-2693(90)91462-K}{\emph{Phys. Lett. B}
  {\bf 242} (1990) 225--233}.

\bibitem{fnw1}
P.~Forgacs, F.~Niedermayer and P.~Weisz, \emph{{The Exact mass gap of the
  Gross-Neveu model. 1. The Thermodynamic Bethe ansatz}},
  \href{http://dx.doi.org/10.1016/0550-3213(91)90044-X}{\emph{Nucl. Phys.} {\bf
  B367} (1991) 123--143}.

\bibitem{mmr-trans}
M.~Mari\~no, R.~Miravitllas and T.~Reis, \emph{{On the structure of
  trans-series in quantum field theory}},
  \href{http://arxiv.org/abs/2302.08363}{{\tt 2302.08363}}.

\bibitem{mmlargeN}
M.~Mari{\~n}o, \emph{{Lectures on non-perturbative effects in large $N$ gauge
  theories, matrix models and strings}},
  \href{http://dx.doi.org/10.1002/prop.201400005}{\emph{Fortsch. Phys.} {\bf
  62} (2014) 455--540}, [\href{http://arxiv.org/abs/1206.6272}{{\tt
  1206.6272}}].

\bibitem{mmbook}
M.~Mari{\~n}o, \emph{Instantons and large $N$. An introduction to
  non-perturbative methods in quantum field theory}.
\newblock Cambridge University Press, 2015.

\bibitem{ss}
T.~M. Seara and D.~Sauzin, \emph{Resumaci\'o de {B}orel i teoria de la
  ressurgencia}, {\emph{Butl. Soc. Catalana Mat.} {\bf 18} (2003) 131--153}.

\bibitem{abs}
I.~Aniceto, G.~Ba\c{s}ar and R.~Schiappa, \emph{A primer on resurgent
  transseries and their asymptotics},
  \href{http://dx.doi.org/10.1016/j.physrep.2019.02.003}{\emph{Phys. Rep.} {\bf
  809} (2019) 1--135}.

\bibitem{nsvz-tr}
V.~A. Novikov, M.~A. Shifman, A.~I. Vainshtein and V.~I. Zakharov,
  \emph{{Calculations in external fields in quantum chromodynamics. Technical
  review}}, \href{http://dx.doi.org/10.1002/prop.19840321102}{\emph{Fortsch.
  Phys.} {\bf 32} (1984) 585--622}.

\bibitem{pascual-tarrach}
P.~Pascual and R.~Tarrach, \emph{{QCD: renormalization for the practicioner}},
  vol.~194 of \emph{Lecture Notes in Physics}.
\newblock Springer-Verlag, 1984.

\bibitem{pascual-der}
P.~Pascual and E.~de~Rafael, \emph{{Gluonic Corrections to Quark Vacuum
  Condensate Contributions to Two Point Functions in {QCD}}},
  \href{http://dx.doi.org/10.1007/BF01548609}{\emph{Z. Phys. C} {\bf 12} (1982)
  127}.

\bibitem{elias}
V.~Elias, \emph{{Quantum Field Theory and the Dynamical Generation of Quark
  Masses}}, \href{http://dx.doi.org/10.1139/p86-110}{\emph{Can. J. Phys.} {\bf
  64} (1986) 595--601}.

\bibitem{huangs}
T.~Huang and Z.~Huang, \emph{{Quantum Chromodynamics in Background Fields}},
  \href{http://dx.doi.org/10.1103/PhysRevD.39.1213}{\emph{Phys. Rev. D} {\bf
  39} (1989) 1213--1220}.

\bibitem{giombi}
S.~Giombi, V.~Kirilin and E.~Skvortsov, \emph{{Notes on Spinning Operators in
  Fermionic CFT}}, \href{http://dx.doi.org/10.1007/JHEP05(2017)041}{\emph{JHEP}
  {\bf 05} (2017) 041}, [\href{http://arxiv.org/abs/1701.06997}{{\tt
  1701.06997}}].

\bibitem{beneke-braun2}
M.~Beneke and V.~M. Braun, \emph{{Heavy quark effective theory beyond
  perturbation theory: Renormalons, the pole mass and the residual mass term}},
  \href{http://dx.doi.org/10.1016/0550-3213(94)90314-X}{\emph{Nucl. Phys. B}
  {\bf 426} (1994) 301--343}, [\href{http://arxiv.org/abs/hep-ph/9402364}{{\tt
  hep-ph/9402364}}].

\bibitem{avan-vega}
J.~Avan and H.~J. de~Vega, \emph{{I}nstantons of two-dimensional fermionic
  effective actions by inverse scattering transformation},
  \href{http://dx.doi.org/10.1007/BF01209295}{\emph{Commun. Math. Phys.} {\bf
  102} (1985) 463}.

\bibitem{dmss}
L.~Di~Pietro, M.~Mari\~no, G.~Sberveglieri and M.~Serone, \emph{{Resurgence and
  1/N Expansion in Integrable Field Theories}},
  \href{http://dx.doi.org/10.1007/JHEP10(2021)166}{\emph{JHEP} {\bf 10} (2021)
  166}, [\href{http://arxiv.org/abs/2108.02647}{{\tt 2108.02647}}].

\bibitem{balog}
J.~Balog, \emph{{Form-factors and asymptotic freedom in the O(3) sigma model}},
  \href{http://dx.doi.org/10.1016/0370-2693(93)90762-7}{\emph{Phys. Lett. B}
  {\bf 300} (1993) 145--151}.

\bibitem{balog-nh}
J.~Balog, M.~Niedermaier and T.~Hauer, \emph{{Perturbative versus
  nonperturbative QFT: Lessons from the O(3) NLS model}},
  \href{http://dx.doi.org/10.1016/0370-2693(96)00906-9}{\emph{Phys. Lett. B}
  {\bf 386} (1996) 224--232}, [\href{http://arxiv.org/abs/hep-th/9604161}{{\tt
  hep-th/9604161}}].

\bibitem{kleinert}
H.~Kleinert, \emph{Particles and quantum fields}.
\newblock World Scientific, 2016.

\end{thebibliography}\endgroup

\end{document}